\documentclass[aps,pra,superscriptaddress,longbibliography,twocolumn]{revtex4-2}

\usepackage{array}
\usepackage{graphicx}
\usepackage{amsmath}
\usepackage{amssymb}
\usepackage{hyperref}
\usepackage[utf8]{inputenc}
\usepackage{mathtools}
\usepackage[english]{babel}
\usepackage{bbm}
\hypersetup{colorlinks=true, linkcolor=blue, citecolor=blue, urlcolor=blue}
\usepackage{xcolor}
\usepackage{braket}
\usepackage{bbold}
\usepackage[normalem]{ulem}
\usepackage{dsfont}

\newcommand{\ie}{i.\,e.,\ }

\newcommand{\eg}{e.\,g.,\ }

\newcommand{\bea}{\begin{eqnarray}}
\newcommand{\eea}{\end{eqnarray}}

\usepackage{pgf}
\usepackage{layouts}

\begin{document}
\title{
Symmetry based efficient simulation of higher-order coherences \\
in quantum many-body superradiance}

\author{Raphael Holzinger}
\email{raphael3d@gmail.com}
\affiliation{Institute for Theoretical Physics, University of Innsbruck, A-6020 Innsbruck, Austria}

\author{Oriol Rubies-Bigorda}
\affiliation{Physics Department, Massachusetts Institute of Technology, Cambridge, Massachusetts 02139, USA}
\affiliation{Department of Physics, Harvard University, Cambridge, Massachusetts 02138, USA}

\author{Susanne F. Yelin}
\affiliation{Department of Physics, Harvard University, Cambridge, Massachusetts 02138, USA}

\author{Helmut Ritsch}
\affiliation{Institute for Theoretical Physics, University of Innsbruck, A-6020 Innsbruck, Austria}

\begin{abstract}
We propose an efficient method to numerically simulate the superradiant emission dynamics of large numbers of quantum emitters in ordered arrays in the presence of long-range dipole-dipole interactions mediated by the vacuum electromagnetic field. Using the spatial symmetries of the system, we rewrite the equations of motion in a collective spin basis and subsequently apply a higher-order cumulant expansion for the collective operators. By truncating the subradiant collective modes with a heavily suppressed decay rate and keeping only the effect from the radiating collective modes, we reduce the numerical complexity significantly. This allows to efficiently compute the dissipative dynamics of the observables of interest for a linear and ring-shaped arrays of quantum emitters. In particular, we characterize the second order intensity correlation function $g^{(2)}(\tau =0)$, which is challenging to compute for extended systems with traditional cumulant expansion methods.
\end{abstract}

\maketitle

\section{Introduction}
\begin{figure}[ht!]
    \centering
    \includegraphics[width = 1\columnwidth]{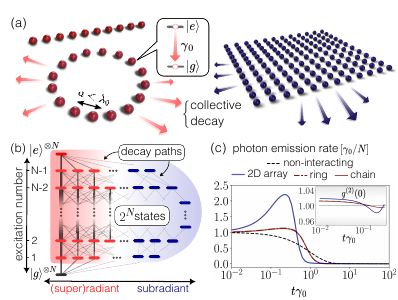}
    \caption{(a) Ordered arrays of $N$ collectively decaying two-level quantum emitters with subwavelength separation $a < \lambda_0= 2\pi c/\omega_0$, where $\lambda_0$ is the transition wavelength. (b) Illustration of the decay dynamics for an initially fully excited ensemble with eigenstates organized according to their excitation number (up to down) and decay rates (left to right). Only a small number of eigenstates are significantly occupied during the time evolution (in red) while a majority of subradiant states (in blue) have a small occupation probability. In this work, we show how to simulate $\sim ( a/\lambda_0) N$ (1D arrays) and $\sim (a/\lambda_0)^2 N$ (2D arrays) collective modes instead of $N$ individual emitters, thus reducing the numerical complexity significantly. (c) The photon emission rate $p_\mathrm{out}(t)$ in Eq.~\ref{p_out} and second order correlation function $g^{(2)}(\tau = 0)$ in Eq.~\ref{eq:g2} at early times, computed by truncating (\ie neglecting) the subradiant collective modes. We show the results for a chain and a ring of $100$ two-level emitters, as well as for a square array of $400$ atoms. We consider circularly polarized emitters with lattice spacing $a/\lambda_0=0.15$.} 
    \label{fig:fig1}
\end{figure}

The accurate description of non-equilibrium dynamics such as collective radiative decay of interacting quantum emitters is one of the major challenges of open system quantum many-body theory. It is of central importance in many areas of physics, since an interacting array of emitters behaves quite differently from its individual constituents. When coupled via the electromagnetic field~\cite{reitz2022cooperative}, excited emitters can trigger collective many-body quantum phenomena, resulting in short and intense bursts of light (generally termed superradiance or superfluorescence~\cite{masson2020many,Ana_superradiance_2, sutherland2017superradiance,rubies2022superradiance,parmee2020signatures}) as well as directional emission of light~\cite{Ana_superradiance_2,holzinger2021nanoscale}, which has been oberserved in numerous experiments~\cite{ferioli2022observation,glicenstein2022superradiance,ferioli2021laser,ferioli2021storage,inouye1999superradiant,grimes2017direct,kaluzny1983observation,chen2018experimental,pellegrino2014observation}. Understanding collective dissipation is crucial for enabling fundamental studies in quantum many-body physics~\cite{Chang2018,HUANG2023100470}, in the development of novel light sources~\cite{holzinger2020nanoscale} or quantum metrology with dense spin ensembles~\cite{cox2016deterministic}. It is also vital to understand its implication on quantum error correction, where collective decay can directly impact fault-tolerance in quantum systems involving many emitters~\cite{cattaneo2023quantum,lemberger2017effect,aharonov2006fault}. Ordered atomic emitter arrays with subwavelength separations have emerged in recent years as a platform to study such dissipative many-body open quantum dynamics involving long-range interactions~\cite{gross_haroche,ferioli2021laser,gold2022spatial}, illustrated in Fig.~\ref{fig:fig1}(a). Physical realizations include alkaline-earth atoms trapped in optical lattices~\cite{gyger2024continuous,norcia2024iterative,patscheider2020controlling,glicenstein2021preparation,hutson2024observation} and in optical tweezer arrays~\cite{barredo2016atom,browaeys2020many,labuhn2016tunable,cooper2018alkaline,norcia2018microscopic}, which allow to trap thousands of atoms in ordered geometries~\cite{manetsch2024tweezer,pause2024supercharged}.
\begin{figure*}[ht!]
    \centering
    \includegraphics[width = 1\textwidth]{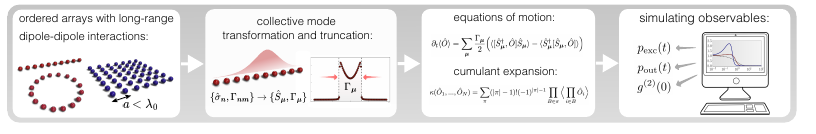}
    \caption{The protocol for simulating dissipative quantum many-body systems using the collective mode truncation presented in this paper. Starting from ordered arrays of quantum emitters with subwavelength separations $a$, shown in Section~\ref{section2}, we transform the system into the collective mode basis and apply the collective mode truncation in Sections~\ref{section2}. Next, we write down the equations of motion for the expectation values in the collective mode basis. For that, we apply a cumulant expansion to obtain a closed set of equations in Section~\ref{sec:eom}. Finally, the time evolution of the expectation values of the observables of interest are numerically simulated. In our case, these are the excited state population $p_\mathrm{ext}(t)$ in Eq.~\ref{p_exc}, the total photon emission rate $p_\mathrm{out}(t)$ in Eq.~\ref{p_out} and the second order correlation function $g^{(2)}(0)$ in Eq.~\ref{eq:g2}, presented in Section~\ref{sec:eom}.}
    \label{fig:fig22}
\end{figure*}

The Hilbert space of an ensemble of $N$ emitters grows exponentially with $N$. In the presence of collective dissipation or other decoherence processes, describing the dynamics of the system requires to solve for the $2^N \times 2^N$ density matrix over time. 
For smaller numbers of emitter, one can utilize a Monte Carlo wavefunction method, which converges to the exact expectation values based on the master equation when averaging over a sufficient number of trajectories~\cite{verstraelen2023quantum}.
Alternatively, for traditional cumulant expansion methods that include quantum correlations beyond a bare mean field treatment ~\cite{kubo1962generalized,rubies2023characterizing,sanchez2020cumulant,masson2024dicke,robicheaux2021beyond,kramer2015generalized, Rubies_subradiance_cumulant}, the number of equations scales polynomially with the number of emitters, rendering the simulation of more than a few hundred emitters challenging. In particular, numerical simulations of higher order correlations (such as the second order intensity correlation function) are challenging, but carry great interest for studying the quantum statistics and coherence properties of the emitted light~\cite{ferioli2021laser,ferioli2022observation}. 

In this work, we present a novel way to calculate the dynamics and the higher order correlation functions in regular arrays by exploiting the symmetric nature of the collective super- and subradiant modes~\cite{Dicke_originalpaper,asenjo2017exponential} that emerge at subwavelength separations between the emitters, as illustrated in Fig.~\ref{fig:fig1}(b). We demonstrate the effectiveness of this method by studying collective decay cascades in large ordered arrays of emitters, for which we show numerical simulations systems containing up to 400 emitters in Fig.~\ref{fig:fig1}(c). Notably, this formalism can be leveraged to study the collective dissipative dynamics of quantum emitters (irrespective of their nature) coupled to a common radiation field. Apart from cold atoms, the formalism can also describe molecules~\cite{wurthner2011j}, collective phenomena in waveguide quantum electrodynamics~\cite{pennetta2022collective,tebbenjohanns2024predicting,liedl2024observation,goban2015superradiance}, magnons in solid-state environments~\cite{li2023solid} and quantum dots in self-assembled superlattices~\cite{raino2018superfluorescence}.
\newline
This work is organized as follows: First, we give a compact introduction to the open quantum system model used to describe the many-body dynamics involving two-level systems that radiatively decay. We introduce the collective mode picture for interacting quantum emitters in ordered arrays and the higher-order cumulant expansion for the Heisenberg equations of motion. With these ingredients, we motivate and propose the collective mode truncation, which is the main result of this work. The full protocol of the method is illustrated in Fig.~\ref{fig:fig22}. Next, we benchmark our method with the exact numerics based on a master equation time evolution with a small number of emitters placed in a ring geometry. We then show the scaling of the total emission rate for large ensembles as a function of $N$. We make use of our truncation method to simulate system sizes that are computationally too costly for traditional cumulant expansion methods . Finally, we summarize the results and give an outlook to extend this method to a broader range of scenarios.  

\section{Collective decay in quantum emitter arrays} \label{section2}
First, we introduce the theoretical framework used to desribe the open quantum system dynamics of ordered arrays emitters interacting through the common free-space electromagnetic environment. We consider $N$ atomic two-level emitters trapped in a periodic array with sub-wavelength nearest-neighbor separation $a$. Each emitter features a resonance frequency $\omega_0 = 2\pi/\lambda_0$ and spontaneous decay rate $\gamma_0=k_0^3 |\boldsymbol{d}|^2 /(3\pi \epsilon_0 \hbar)$, where $\boldsymbol{d}$ is the transition dipole moment and $\epsilon_0$ the vacuum permittivity. 
For simplicity we assume an angular momentum $J=0$ ($\ket{g}$) to $J'=1$ ($\ket{e}$) transition with an unique ground state ($\ket{g}$). 
After tracing out the vacuum electromagnetic field modes under the Born-Markov approximation~\cite{lehmberg1970radiation,novotny2012principles}, the dynamics of the system's density matrix $\hat{\rho} = |\psi \rangle \langle \psi|$ is governed by a master equation of the form ($\hbar=1$)
\begin{equation} \label{master}
    \frac{d}{dt} \hat{\rho} = -i [\hat{H},\hat{\rho}] + \sum_{{n,m}}^N \frac{\Gamma_{{{nm}}}}{2} \Big(2  \hat{\sigma}_{{n}} \hat{\rho} \hat{\sigma}_{{m}}^\dagger  -  \{ \hat{\sigma}^\dagger_{{n}}  \hat{\sigma}_{{m}} ,\hat{\rho} \}  \Big),
\end{equation}
where the $\hat{H}= \sum_n \omega_0 \hat{\sigma}^{ee}_n + \sum_{{n} \neq {m}} J_{{nm}} \hat{\sigma}^\dagger_{{n}} \hat{\sigma}_{{m}}$ desribes coherent dynamics between emitters. The operators $\hat{\sigma}_{{n}} = |g_{{n}} \rangle \langle e_{{n}}|$ act as the transition operators for emitters at positions $\{ \mathbf{r}_n \}$ and $\hat{\sigma}^{ee}_n \equiv \hat{\sigma}^\dagger_n \hat{\sigma}_n$. For the sake of simplicity, we assume identical transition frequencies $\omega_0$ for all emitters throughout this work. The coherent $J_{nm}$ and dissipative $\Gamma_{nm}$ couplings between emitters $n$ and $m$ are related to the Green's tensor in free-space $\mathbf{G}$ (detailed in Appendix~\ref{SI: Green}) via~\cite{asenjo2017exponential}
\begin{align} \label{dipole-interaction}
   \frac{3 \pi \gamma_0}{\omega_0} \mathbf{d}^\dagger \cdot \mathbf{G}(\mathbf{r}_{nm},\omega_0) \cdot \mathbf{d} = -J_{nm} + \frac{i\Gamma_{nm}}{2} ,
\end{align}
where the transition dipole moment $\mathbf{d}$ is assumed to be identical for all emitters and $\mathbf{r}_{nm} = \mathbf{r}_n-\mathbf{r}_m$ is the connecting vector between emitters $n$ and $m$.
At subwavelength separations, $a < \lambda_0=2\pi c/\omega_0$, the virtual photon exchange between emitters induces long-range dipole-dipole interactions $\Gamma_{nm}$. Then, the decay becomes collective and is governed by the jump operators obtained by diagonalizing the decay matrix $\boldsymbol{\Gamma}$ with elements $\Gamma_{nm}$. During a decay cascade, the state of the system can generally pass through a large fraction of the full Hilbert space (which has dimension $2^N$), as illustrated in Fig.~\ref{fig:fig1}(b). Note that for an independent atomic ensemble only individual (diagonal) decay remains, namely $\Gamma_{nm} = \gamma_0 \delta_{nm}$.

\begin{figure*}[ht!]
    \centering
    \includegraphics[width = 1\textwidth]{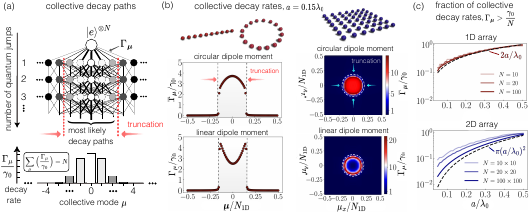}
    \caption{(a) Repeated actions of collective jump operators $\hat{S}_\mu$ (circles) lead to collective decay paths (connecting lines). The most likely decay paths involve (super)radiant jumps with rates $\Gamma_\mu \gtrsim \gamma_0$, while paths generated by subradiant ($\Gamma_{\boldsymbol{\mu}} \ll \gamma_0$) jump operators are unlikely. (b) Collective decay rates for one- and two-dimensional arrays with the dashed lines indicating the border between radiant and subradiant (dark) modes and $a=0.15\lambda_0$. The number of radiant decay channels is strongly reduced at subwavelength separation, comprising a tiny fraction of the total number of $N$ decay channels. In the limit of large $N$, this fraction is given by $2a/\lambda_0$ for one-dimensional arrays and $\pi (a/\lambda_0)^2$ for two-dimensional arrays, as shown in (c).
    In the collective mode picture, we can truncate the system at the dashed lines. (c) Fraction of collective modes with decay rates $\Gamma_{\boldsymbol{\mu}}> \gamma_0/N$ as a function of the array spacing $a$, obtained by numerical diagonalization of the symmetric $N \times N$ decay matrix with elements $\Gamma_{mn}$. For large enough emitter numbers, the fraction of radiant modes follows a linear scaling for one-dimensional arrays and a quadratic scaling for square arrays, leading to a substantial reduction in radiant modes for $a< \lambda_0$.}
    \label{fig:fig2}
\end{figure*}

\subsection*{Collective quantum jumps}

\subsubsection*{Ensemble with translational symmetry}

Let us first consider ensembles of emitters that exhibit permutational symmetry, such as rings with perpendicular or in-plane circular polarization and infinite ordered arrays. Then, the $N$ eigenstates of the dissipative and coherent interactions given by the matrices $\boldsymbol{\Gamma}$ and $\boldsymbol{J}$ (with elements $\Gamma_{nm}$ and $J_{nm}$, respectively) are spin-wave or Bloch operators of the form  
\begin{align} \label{collective-spin}
\hat{S}_{\boldsymbol{\mu}} &= \frac{1}{\sqrt{N}} \sum_{{n=1}}^N\mathrm{exp}\Big(i \frac{2\pi}{N_\mathrm{1D} a}  \boldsymbol{\mu} \cdot \boldsymbol{r}_n  \Big) \ \hat{\sigma}_{{n}}.
\end{align}
The spin waves are labelled by their quasi-momentum. For two-dimensional geometries such as squared arrays with spacing $a$, the quasi-momentum in the $xy$-plane reads $\boldsymbol{k} = 2 \pi \boldsymbol{\mu}/N_{1D}a$, where $\mu_i = 0, \pm 1, \pm 2,\cdots,\lceil \pm(N_\mathrm{1D}-1)/2\rceil$~\cite{asenjo2017exponential}. Here, $N_{1D}$ denotes the number of emitters along one direction, such that the total number of emitters is $N=N_{1D}^2$ for two-dimensional and $N=N_{1D}$ for one-dimensional arrays. For one-dimensional arrys we simply set $\mu_y = 0$. All spin waves are fully delocalized over all sites, and their associated phase profile is dictated by the quasimomentum $\boldsymbol{\mu}$. The collective decay rates and energy shifts associated to each spin wave correspond to the eigenvalues of $\boldsymbol{\Gamma}$ and $\boldsymbol{J}$,
\begin{subequations}\label{collective-rates}
\begin{align} 
    \Gamma_{\boldsymbol{\mu}} &=  \frac{1}{N} \sum_{{n},{m}}^N \mathrm{exp}\Big( i\frac{2\pi}{N_\mathrm{1D} a}  \boldsymbol{\mu} \cdot (\boldsymbol{r}_n-\boldsymbol{r}_m)  \Big) \ \Gamma_{{n} {m}}, \label{decay_rate_PS} \\
    J_{\boldsymbol{\mu}} &= \frac{1}{N} \sum_{{n}\neq{m}}^N \mathrm{exp}\Big( i\frac{2\pi}{N_\mathrm{1D} a}  \boldsymbol{\mu} \cdot (\boldsymbol{r}_n-\boldsymbol{r}_m)  \Big) \ J_{{n} {m}},
    \end{align}
\end{subequations}
with $ \sum_{\boldsymbol{\mu}} \Gamma_{\boldsymbol{\mu}} = N \gamma_0$. From Eq.~\ref{collective-rates} it can be seen that the eigenvalues are symmetric under the inversion $\boldsymbol{\mu} \leftrightarrow -\boldsymbol{\mu}$, meanwhile the collective mode with $\boldsymbol{\mu} = 0$ is always unique.

Notably, the master equation~\ref{master} becomes diagonal in the collective spin basis given by the operators $\hat{S}_{\boldsymbol{\mu}}$, 
\begin{equation} \label{master-coll}
    \frac{d}{dt} \hat{\rho} = -i [\hat{H},\hat{\rho}] + \sum_{\boldsymbol{\mu}}^N \frac{\Gamma_{{\boldsymbol{\mu}}}}{2} \Big(2  \hat{S}_{\boldsymbol{\mu}} \hat{\rho} \hat{S}^\dagger_{\boldsymbol{\mu}}  -  \{ \hat{S}^\dagger_{\boldsymbol{\mu}}  \hat{S}_{\boldsymbol{\mu}} ,\hat{\rho} \}  \Big),
\end{equation}
where the coherent interactions are given by the Hamiltonian $\hat{H}=\sum_{\boldsymbol{\mu}} (\omega_0+J_{\boldsymbol{\mu}}) \hat{S}^\dagger_{\boldsymbol{\mu}} \hat{S}_{\boldsymbol{\mu}}$.

\subsubsection*{General ensembles}

While the collective spin states $\hat{S}_{\boldsymbol{\mu}}$ given in Eq.~(\ref{collective-spin}) are the eigenstates of the coherent and dissipative interactions $\boldsymbol{\Gamma}$ and $\boldsymbol{J}$ for permutationally symmetric ensembles of emitters, this is no longer the case once this symmetry is broken~\cite{asenjo2017exponential,holzinger2020nanoscale,moreno2019subradiance}. For finite arrays, Eq.~(\ref{collective-spin}) are only approximately the eigenstates of the system, and the master equation~(\ref{master-coll}) consequently acquires couplings between spin waves with different quasi-momenta. As shown in Appendix~\ref{appendix:finite} by comparing the decay rates $\Gamma_\nu$ given in Eq.~(\ref{decay_rate_PS}) with the decay rates obtained by diagonalizing the matrix $\boldsymbol{\Gamma}$ exactly, excellent agreement is found specially for large arrays. As a result, the master equation~(\ref{master-coll}) derived for the permutationally symmetric case can be used to simulate and study the dynamics of finite-sized arrays with high accuracy, as discussed in detail in Section~\ref{sec:eom}.

\subsubsection*{Hierarchy of timescales}

In sub-wavlength configurations, the set of collective decay rates $\{ \Gamma_{\boldsymbol{\mu}} \}$ exhibits a high variability and range from superradiant ($\Gamma_{\boldsymbol{\mu}}>\gamma_0$) to strongly subradiant ($\Gamma_{\boldsymbol{\mu}} \ll \gamma_0$). The dissipative dynamics is governed by the set of collective jump operators and decay rates, $\{ \hat{S}_{\boldsymbol{\mu}},\Gamma_{\boldsymbol{\mu}} \}$, and results in a complex network of decay paths from the fully excited state, as sketched in Fig.~\ref{fig:fig2}(a). Notably, a majority of these paths involve subradiant modes at subwavelength separations. These paths occur with small probabilities proportional to their corresponding subradiant decay rates, and can therefore be neglected if one is only interested in the radiant dynamics. intuitively, these occurs due to a separation of timescales, whereby the radiating collective operators act at a much faster rate than their subradiant counterparts. For one-dimensional arrays, Fig.~\ref{fig:fig2}(b) shows that all radiant modes are confined within the lightcone,  $|\mu_{x(y)}|\lesssim a N_\mathrm{1D}$~\cite{moreno2019subradiance,asenjo2017exponential}, leading to a fration of radiant modes equal to $2a/\lambda_0$. In Fig.~\ref{fig:fig2}(c), we numerically show that the number of modes that satisfy $\Gamma_{\boldsymbol{\mu}}>\gamma_0/N$ approaches the above limit as $N$ increases. In the case of two-dimensional square arrays, the light cone is a circle of radius  $a N_\mathrm{1D}$, and the radiant mode fraction for large arrays tends to $ \pi(a/\lambda_0)^2 $, as shown in Fig.~\ref{fig:fig2}(c) as well. As we further discuss in Sections~\ref{sec:eom} and~\ref{sec:results}, neglecting the collective subradiant operators allows to drastically reduce the number of equations that needs to be solved to simulate the superradiant decay dynamics.

\subsection*{Collective photon emission}
In this section, we focus on characterizing the collective photon emission properties of the atomic emitter array. We consider the excited-state population over time
\begin{align} \label{p_exc}
    p_\mathrm{exc}(t) &= \sum_{n=1}^N \langle \hat{\sigma}_n^{ee} \rangle =  N \langle \hat{S}^{ee}_{\boldsymbol{0}} \rangle,
\end{align}
where $\hat{S}^{ee}_{\boldsymbol{0}}$ is the collective excited state operator of the whole array with quasi-momentum  $\boldsymbol{0}=(0,0)$, and reads
\begin{align} \label{collective-ee}
\hat{S}_{\boldsymbol{\mu}}^{ee} &= \frac{1}{{N}} \sum_{{n=1}}^N\mathrm{exp}\Big(-i \frac{2\pi}{N_\mathrm{1D}a}  \boldsymbol{\mu} \cdot \boldsymbol{r}_n  \Big) \ \hat{\sigma}_{{n}}^{ee}.
\end{align}

Next, the total photon emission rate of the array can be written in the individual and collective basis as
\begin{align} \label{p_out}
    p_\mathrm{out}(t) &= \gamma_0 \sum_{n=1}^N \langle \hat{\sigma}^{ee}_n\rangle +  \sum_{n\neq m}^N \Gamma_{nm} \langle \hat{\sigma}_n^{\dagger} \hat{\sigma}_m \rangle \nonumber \\
    &= \sum_{\boldsymbol{\mu}} \Gamma_{\boldsymbol{\mu}} \langle \hat{S}^{\dagger}_{\boldsymbol{\mu}} \hat{S}_{\boldsymbol{\mu}} \rangle,
\end{align}
where we have used Eq.~\ref{collective-spin}. In non-interacting emitters, only the first term in the first line of Eq.~\ref{p_out} is non-zero and exponential decay follows. At subwavelength separations, the light-induced dipole-dipole interactions can resilt in non-zero correlations $\langle \hat{\sigma}_n^{\dagger} \hat{\sigma}_m \rangle$ that extend over the whole array. When these correlations contribute positively, \ie $\Gamma_{nm} \langle \hat{\sigma}_n^{\dagger} \hat{\sigma}_m \rangle > 0$, the total emission rate can attain values $p_\mathrm{out}(t) > \gamma_0 N$ at $t>0$, thereby giving rise to superradiant decay~\cite{rubies2022superradiance}.
\newline
\indent Second order correlations between the atomic emitters allow to characterize two-photon emission statistics of the array during the dissipative dynamics. This is captured by the normalized second order correlation function with zero time delay $g^{(2)}(\tau=0)$~\cite{Ana_superradiance_2,masson2024dicke}, and is expressed both in the individual and collective basis as 
\begin{align} \label{eq:g2}
    g^{(2)}(\tau=0) &=     \frac{\sum_{k,l,m,n}^N \Gamma_{kn} \Gamma_{lm} \langle \hat{\sigma}^{\dagger}_{k} \hat{\sigma}^{\dagger}_{l} \hat{\sigma}_{m} \hat{\sigma}_{n} \rangle}{\Big(\sum_{n,m}^N \Gamma_{nm} \langle \hat{\sigma}^{\dagger}_{m} \hat{\sigma}_{n} \rangle \Big)^2} \nonumber \\
    &=  \frac{\sum_{{\boldsymbol{\mu}} {\boldsymbol{\nu}}} \Gamma_{\boldsymbol{\mu}} \Gamma_{\boldsymbol{\nu}} \langle \hat{S}^{\dagger}_{\boldsymbol{\mu}} \hat{S}^{\dagger}_{\boldsymbol{\nu}} \hat{S}_{\boldsymbol{\nu}} \hat{S}_{\boldsymbol{\mu}} \rangle}{\Big(\sum_{\boldsymbol{\mu}} \Gamma_{\boldsymbol{\mu}} \langle \hat{S}^{\dagger}_{\boldsymbol{\mu}} \hat{S}_{\boldsymbol{\mu}} \rangle \Big)^2}.
\end{align}
Eq.~\ref{eq:g2} is equivalent to the second-order intensity correlation of the total emitted light, detected in the far field, and as such quantifies the emitted photon statistics of the array as a whole. The intensity correlations at any point in space, can be calculated via the electric field operators~\cite{masson2020many,holzinger2021nanoscale} and are discussed in the Appendix~\ref{supp:directional}.
\section{Equations of motion} \label{sec:eom}
In this section, a closed set of equations of motion for expectation values of collective spin operators is derived.
We analyze dissipative dynamics in the absence of external driving, with the system initially in the fully inverted product state~\cite{rubies2023characterizing,robicheaux2021beyond,masson2024dicke},
\begin{equation} \label{initial-state}
   |\psi_0 \rangle = \Big(\prod_{n=1}^N \hat{\sigma}^\dagger_n\Big)  |g\rangle^{\otimes N}= |e\rangle^{\otimes N}.
\end{equation}
In terms of density matrices, this product state can be written as $\hat{\rho}(0) = \otimes_{n=1}^N \hat{\rho}_n$, with $\hat{\rho}_n = |e_n \rangle \langle e_n|$. 
To numerically simulate the time evolution of operator expectation values $\langle \hat{O} \rangle$, one can use the Master equation in Eq.~\ref{master} and evaluate $\frac{d}{dt} \langle \hat{O} \rangle = \mathrm{Tr}(\hat{O}  \frac{d}{dt}\hat{\rho})$. Equivalently, the Heisenberg equations of motion can be considered to express the time evolution for expectation values as~\cite{robicheaux2021beyond,rubies2023characterizing,sanchez2020cumulant}
\begin{align} \label{heisenberg-collective}
    \frac{d}{dt} \langle \hat{O} \rangle &=  i \langle [\hat{H},\hat{O}]\rangle \nonumber \\
    &-\sum_{\boldsymbol{\mu}} \frac{\Gamma_{\boldsymbol{\mu}}}{2} \Big( \langle [\hat{S}^\dagger_{\boldsymbol{\mu}},\hat{O}] \hat{S}_{\boldsymbol{\mu}} \rangle -  \langle \hat{S}^\dagger_{\boldsymbol{\mu}} [\hat{S}_{\boldsymbol{\mu}},\hat{O}]\rangle
    \Big),
\end{align}
where we neglected quantum noise terms~\cite{Lehmberg1970_1,rubies2023characterizing}, as we are interested in operator averages. The operator $\hat{O}$ is generally a product involving $r$ operators $\hat{O}_1 \hat{O}_2 \cdots\hat{O}_r $. In most instances, the time derivative of $\langle \hat{O} \rangle$ will depend on expectation values of products of up to $r+1$ operators, on the right hand side of Eq.~\ref{heisenberg-collective}. This will lead to a growing system of equations whose size scales polynomially with $N$. In order to obtain a closed set of equations, certain higher-order correlations between operators have to be neglected, which is achieved systematically through the cumulant expansion~\cite{Cumulant_Kubo,Robicheaux_cumulants}. 
Appendix~\ref{supp:cumulant-collective} provides a detailed discussion of the cumulant expansion and the resulting equations of motion, while we focus here on a compact description.
Before proceeding, we would like to highlight that collective spin operators with different quasi-momenta $\boldsymbol{\mu} \neq \boldsymbol{\nu}$ do not commute in general, $ [\hat{S}^\dagger_{\boldsymbol{\mu}},\hat{S}_{\boldsymbol{\nu}}] \neq 0$. This is in stark contrast to the local spin basis, for which opperators associated to different spins always commute, \eg $[\hat{\sigma}^\dagger_n,\hat{\sigma}_m] = (2\hat{\sigma}_{n}^{ee} -\hat{\mathds{1}} ) \delta_{nm}$ commutes for $n\neq m$. This fact leads to couplings among modes of different quasi-momenta in the equations of motion and induces complex decay paths en route to the ground state $|g\rangle ^{\otimes N}$ as is illustrated in Fig.~\ref{fig:fig1}(b) and Fig.~\ref{fig:fig2}(a). The set of commutation relations read
\begin{subequations} \label{commutation1}
\begin{align} 
\Big[\hat{S}_{\boldsymbol{\mu}}^\dagger,\hat{S}_{\boldsymbol{\nu}} \Big] &= 2 \hat{S}_{\boldsymbol{\mu}-\boldsymbol{\nu}}^{ee} - \hat{\mathds{1}} \delta_{_{\boldsymbol{\mu \nu}}}, \\
\Big[\hat{S}^\dagger_{\boldsymbol{\mu}},\hat{S}_{\boldsymbol{\nu}}^{ee} \Big]  &= - \frac{1}{N} \hat{S}^\dagger_{\boldsymbol{\mu+\nu}}, \\
\Big[\hat{S}_{\boldsymbol{\mu}},\hat{S}_{\boldsymbol{\nu}}^{ee} \Big]  &= \frac{1}{N} \hat{S}_{\boldsymbol{\mu-\nu}},
\end{align}
\end{subequations}
 where $\hat{\mathds{1}}$ is the identity operator. 
 In Eq.~\ref{commutation1}(a) we have used the orthogonality of the sum for periodic boundary conditions
 \begin{equation} \label{ortho}
     \sum_{n=1}^N \mathrm{exp} \Big( i \frac{2\pi}{N_\mathrm{1D}a} {\boldsymbol{\mu}} \cdot \boldsymbol{r}_n \Big) = N\delta_{ \boldsymbol{\mu 0}},
 \end{equation}
 with $N = N_\mathrm{1D}^2$ for two-dimensional and $N=N_\mathrm{1D}$ for one-dimensional arrays.

In order to obtain the equations of motion for the expectation values of the collective spin operators, we start from the set of operators $\{\hat{S}^\dagger_{\boldsymbol{\mu}},\hat{S}_{\boldsymbol{\mu}},\hat{S}^{ee}_{\boldsymbol{\mu}} \}$ and apply Eq.~\ref{heisenberg-collective} to derive their dynamics in terms of higher-order operators. In turn we repeat this procedure for the higher-order operators, too. One important consequence of the phase symmetry of the initial product state in Eq.~\ref{initial-state} is the fact that expectation values such as $\langle \hat{S}_{\boldsymbol{\mu}}\rangle$ and $\langle \hat{S}_{\boldsymbol{\mu}} \hat{S}_{\boldsymbol{\nu}}^{ee} \rangle$ are zero throughout the whole time evolution~\cite{masson2024dicke,rubies2023characterizing}. This also generates a constraint on the sum of quasi-momenta for the expecation values. Additionally, the initial state in Eq.~\ref{initial-state} also implies $\langle \hat{{S}}_{\boldsymbol{\mu}}^{\dagger} \hat{{S}}_{\boldsymbol{\nu}} \rangle = \delta_{\boldsymbol{\mu} \boldsymbol{\nu}}$, where we have used Eq.~\ref{ortho}. Indeed, one can numerically check that only expectation values of the form $\langle \hat{{S}}_{\boldsymbol{\mu}}^{\dagger} \hat{{S}}_{\boldsymbol{\mu}} \rangle$ are non-zero throughout the time dynamics for ring geometries (for which Eq.~\ref{ortho} is exact). This results in a significant reduction of the number of expectation values and the equations of motion that need to be simulated over time.
Starting from the Heisenberg equations of motion in Eq.~\ref{heisenberg-collective}, we finally find the equations of motion for the following set of expecation values
\begin{align} \label{eq:collective}
    \Big\{  &\langle \hat{{S}}_{\boldsymbol{0}}^{ee} \rangle ,  \langle \hat{{S}}_{\boldsymbol{\mu}}^{\dagger} \hat{{S}}_{\boldsymbol{\mu}} \rangle,  \langle \hat{{S}}_{\boldsymbol{\mu}}^{ee} \hat{{S}}_{-\boldsymbol{\mu}}^{ee} \rangle,  \\
    &\langle \hat{{S}}^\dagger_{\boldsymbol{\mu}} \hat{{S}}_{\boldsymbol{\nu}} \hat{{S}}_{\boldsymbol{\nu}-\boldsymbol{\mu}}^{ee} \rangle,  \langle \hat{\mathcal{S}}^\dagger_{\boldsymbol{\mu}} \hat{\mathcal{S}}^\dagger_{\boldsymbol{\nu}} \hat{\mathcal{S}}_{\boldsymbol{\xi}} \hat{\mathcal{S}}_{\boldsymbol{\nu}-\boldsymbol{\mu}-\boldsymbol{\xi}} \rangle 
    \Big\} \nonumber,
\end{align}
which in turn couple to expectation values such as $\langle \hat{{S}}^\dagger_{\boldsymbol{\mu}} \hat{{S}}_{\boldsymbol{\nu}}^\dagger \hat{{S}}_{\boldsymbol{\xi}} \hat{{S}}_{\boldsymbol{o}} \hat{{S}}_{\boldsymbol{\xi}+\boldsymbol{o}-\boldsymbol{\mu}-\boldsymbol{\nu}}^{ee} \rangle$ and
$\langle \hat{\mathcal{S}}^\dagger_{\boldsymbol{\mu}} \hat{\mathcal{S}}_{\boldsymbol{\nu}} \hat{\mathcal{S}}_{\boldsymbol{\xi}}^{ee} \hat{\mathcal{S}}_{\boldsymbol{\nu}-\boldsymbol{\mu}-\boldsymbol{\xi}}^{ee} \rangle$. These terms are expanded in terms of lower order expectation values using the cumulant expansion method. The explicit expressions for the equations of motions as well as the cumulant expansions are provided in Appendix~\ref{supp:cumulant-collective}. 

The values for the expectation values in Eq.~\ref{eq:collective} can be evaluated at $t=0$ and we obtain
\begin{subequations} \label{eom}
\begin{align}
    \langle \hat{S}^{ee}_{\boldsymbol{\mu}}\rangle&= \delta_{\boldsymbol{\mu} \boldsymbol{0}}, \\
\langle \hat{S}^{\dagger}_{\boldsymbol{\mu}} \hat{S}_{\boldsymbol{\mu}} \rangle&= 1 \ \mathrm{for \ all \ } \boldsymbol{\mu}, \\
\langle \hat{S}^{ee}_{\boldsymbol{\mu}} \hat{S}^{ee}_{-\boldsymbol{\mu}} \rangle&= \delta_{{\boldsymbol{\mu}} \boldsymbol{0}}, \\
\langle \hat{S}^{\dagger}_{\boldsymbol{\mu}} \hat{S}_{\boldsymbol{\nu}} \hat{S}^{ee}_{\boldsymbol{\nu}-\boldsymbol{\mu}} \rangle&=  \delta_{\boldsymbol{\mu} \boldsymbol{\nu}}, \\
\langle \hat{S}^{\dagger}_{\boldsymbol{\mu}} \hat{S}^{\dagger}_{\boldsymbol{\nu}} \hat{S}_{\boldsymbol{\xi}} \hat{S}_{\boldsymbol{\mu}+\boldsymbol{\nu}-\boldsymbol{\xi}} \rangle&= -2/N +\delta_{\boldsymbol{\mu} \boldsymbol{\xi}}+\delta_{\boldsymbol{\nu} \boldsymbol{\xi}}.
\end{align}
\label{initial}
\end{subequations}
Note that the expectation values in Eq.~\ref{eom}(e) necessary to evaluate the second order correlation $g^{(2)}(0)$ involve products of four operators, but only contain three momentum indices $\boldsymbol{\mu},\boldsymbol{\nu},\boldsymbol{\xi}$. This results in $\mathcal{O}(N^3)$ terms. Conversly, the individual spin basis threquires to solve for $\mathcal{O}(N^4)$ terms of the form $\langle \hat{\sigma}^\dagger_n \hat{\sigma}_m^\dagger \hat{\sigma}_k \hat{\sigma}_l \rangle$, making the evalution of the $g^{(2)}(0)$-function numerically challenging for large $N$.
\subsection*{Collective mode truncation}
The major step in order to reduce the computational cost in solving the time evolution for the set of expectation values in Eqs.~\ref{eq:collective} consists in reducing the number of collective modes considered. We illustrate this for one-dimensional arrays ($N=N_\mathrm{1D}$) with quasi-momentum $\mu$, athough these concepts can be readily extended to two-dimensional arrays with quasi-momenta $\boldsymbol{\mu}=(\mu_x,\mu_y)$.
    In the limit of large $N$, as shown in Fig.~\ref{fig:fig2} (a) and (b), there are $\tilde{N}$ (super)radiant modes that satisfy $|{\mu}| \leq a N$. If one is interested in the radiating part of the the dissipative time dynamics, one can keep these radiating modes while discarding the subradiant ones.
For finite $N$, we numerically compute the collective decay rates in Eq.~\ref{collective-rates} and determine the number of modes $\tilde{N}$ satisfying $\Gamma_\mu > \gamma_0/N$.
This leads to a reduction of the number of collective modes to  $\mu = 0, \pm 1, \pm 2,\cdots,\lceil \pm(\tilde{N}-1)/2\rceil$. The scaling of the fraction of modes satisfying this condition as a function of the array spacing $a$ is shown in Fig.~\ref{fig:fig2}(c) for both one- and two-dimensional arrays.

As we show in Fig.~\ref{fig:fig2}(c), the number of radiant modes scales as $2a/\lambda_0 N$ and $\pi (a/\lambda_0)^2 N$ for one- and two-dimensional arrays respectively, which leads to a substantial reduction of the relevant number of modes for arrays with subwavelength spacings between nearest neighbors. We list a comparison of the number of equations necessary to solve the time dynamics in Eq.~\ref{master} and Eq.~\ref{heisenberg-collective} between different numerical methods in Table~\ref{scaling} in terms.  
Moreover, the second-order correlation function $g^{(2)}(0)$ in the individual spin basis involves expectation values of products of four operators and thus $\mathcal{O}(N^4)$ equations need to be solved to obtain its time evolution. However, in the collective spin basis this reduces to $\mathcal{O}(N^3)$, as the necessary expectation values $\langle \hat{\mathcal{S}}^\dagger_{\boldsymbol{\mu}} \hat{\mathcal{S}}^\dagger_{\boldsymbol{\nu}} \hat{\mathcal{S}}_{\boldsymbol{\xi}} \hat{\mathcal{S}}_{\boldsymbol{\nu}-\boldsymbol{\mu}-\boldsymbol{\xi}} \rangle$ only involve three quasi-momenta. This allows to evaluate the time evolution for the $g^{(2)}(0)$ function for large numbers of emitters, as will be shown in the next section.

\begin{table}
\begin{center}
\begin{tabular}{ | m{15.4em} | m{9.5em}| } 
  \hline
    \vspace{0.25em}
  \hfil Method & \hfil Number of equations  \\ 
  \hline\hline
  \vspace{0.5em}
  Quantum jump method (MCWF) &  \hfil $N_\mathrm{traj} \cdot 2^N$ \\ 
  \hline
  \vspace{0.5em}
  Cumulant expansion 4th order&  \hfil $  \sim N^{4} $ \\ 
  \hline 
  \vspace{0.75em}
   Collective mode truncation in 1D &  \hfil $   \sim N^3 \cdot (a/\lambda_0 )^3   $ \\
  \hline
    \vspace{0.75em}
   Collective mode truncation in 2D &  \hfil $  \sim N^3 \cdot  (a/\lambda_0 )^6  $   \\ 
  \hline
\end{tabular}
\caption{The numerical complexity, in terms of the number of equations for different methods that are required to solve the dissipative quantum dynamics in order to evaluate $g^{(2)}(\tau=0)$ in Eq.~\ref{eq:g2}. The quantum jump method scales exponentially with the number of emitters $N$ and has to be averaged over $N_\mathrm{traj}$ trajectories~\cite{dum1992monte,molmer_monte_1993}. The cumulant expansion scales polynomially with the number of emitters~\cite{Cumulant_Kubo}. For the collective mode truncation meanwhile, the numerical complexity is reduced by orders of magnitude for subwavelength spacings $a<\lambda_0$ between nearest neighbors.}
\label{scaling}
\end{center}
\end{table}
\begin{figure}[ht!]
    \centering
    \includegraphics[width = 1\columnwidth]{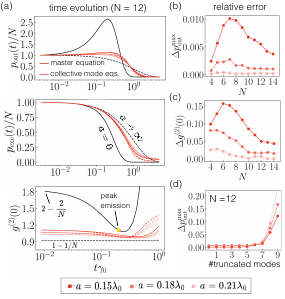}
    \caption{Benchmarking the time evoluation of the expectation values in Eq.~\ref{eq:collective} (dashed lines) against exact numerics using the master equation in Eq.~\ref{master} (continuous lines) for emitters in a ring geometry. (a) The total photon emission $p_\mathrm{out}(t)$ (Eq.~\ref{p_out}), the excited state population $p_\mathrm{ex}(t)$ (Eq.~\ref{p_exc}) and the second order correlation function $g^{(2)}(\tau=0)$ (Eq.~\ref{eq:g2}) as a function of time. The black continuous and dashed curves show the limiting cases $a = 0 $ in the absence of coherent couplings $J_\mathrm{nm}$ (Dicke superradiance) and non-interacting emitters with $a \rightarrow \infty$, respectively. (b) Relative error $\Delta p^{\mathrm{max}}_\mathrm{out}$ of the peak photon emission rate as a function of the emitter number $N$. (c) Relative error $\Delta g^{(2)}(0)$ of the second order correlation function at peak photon emission. In (a), (b) and (c), no modes are truncated when solving the cumulant expansion in the collective mode basis. (d) Relative error of the peak photon emission as a function of the number of truncated modes. For $N=12$ emitters, up to seven modes ($\sim 60 \%$) can be neglected before the relative error increases significantly. We consider $\boldsymbol{d} = (1,i,0)^T/\sqrt{2}$ in all plots.}
    \label{fig:fig4}
\end{figure}
\section{Results and Benchmarking} \label{sec:results}
In this section, we present the application of the cumulant expansion in the collective mode basis and the collective mode truncation for specific geometries. We first consider emitters in a ring geometry, for which the collective mode transformation in Eq.~\ref{collective-spin} is exact independently of $N$. The ring geometry can therefore be used to benchmark the validity of the method with exact numerics based on the time evolution of the density matrix in Eq.~\ref{master}.
In the second part of this section, we apply the collective mode truncation to simulate one- and two-dimensional arrays with large numbers of emitters and show a comparison with the traditional cumulant expansion for individual spin operators~\cite{robicheaux2021beyond,rubies2023characterizing,masson2024dicke}. In Appendix~\ref{appendix:finite} we discuss the extensions to three-dimensional cubic arrays and arrays coupled to one-dimensional waveguides.

The numerical simulations are performed within the Julia programming language~\cite{bezanson2017julia} and the time dynamics are computed using the DifferentialEquations.jl~\cite{rackauckas2017differentialequations} package.
\subsection*{Comparison with exact numerics}
Here, we consider emitters in a ring geometry and provide a comparison with exact numerics based on the master equation. In Fig.~\ref{fig:fig4}(a), we show the time evolution for the observables introduced in Section~\ref{section2} for $12$ emitters with circular in-plance polarization, $\boldsymbol{d}=(1,i,0)^T/\sqrt{2}$ and for various spacings $a$. The collective mode basis shows excellent agreement with the master equation for the photon emission rate $p_\mathrm{out}(t)$ as well as the excited state population $p_\mathrm{ex}(t)$. In both cases, non-interacting emitters with exponential decay $ \mathrm{exp}(-t \gamma_0)$ are shown as well. We plot the relative error $\Delta p_\mathrm{out}^\mathrm{max}$ (between the collective mode basis and the master equation) as a function of $N$ in Fig.~\ref{fig:fig4}(b) for the photon emission rate at peak emission, and observe a reduction of the relative error with increasing $N$.
\begin{figure}[ht!]
    \centering
\includegraphics[width=0.95\columnwidth]{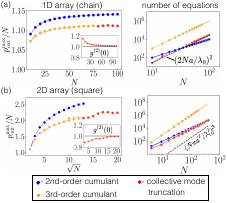}
    \caption{Comparison of the peak photon emission and numerical complexity as a function of $N$ using a second- and third order cumulant expansion in the individual spin basis and the cumulant expansion in the collective spin basis with collective mode truncation. The linear chain in (a) and the square array in (b) show excellent agreement between the third order cumulant expansion and the collective mode equations in Eq.~\ref{eq:collective}. Here, all subradiant modes are truncated and the scaling of the numerical complexity is shown on the right-hand side (red dots). The two-dimensional square array exhibits small oscillations, stemming from the collective mode truncation. The second order correlation $g^{(2)}(0)$ at peak photon emission, evaluated with the collective mode truncation shows a convergence toward unity at peak photon emission for increasing $N$. The second order cumulant expansion clearly overestimates the peak photon emission rate both for chain and square geometries~\cite{rubies2023characterizing,masson2024dicke}. We consider $a=0.15\lambda_0$ and $\mathbf{d}=(1,i,0)^T/\sqrt{2}$ in all plots.}
    \label{fig:fig5}
\end{figure}

The second order correlation function $g^{(2)}(0)$ is a key and very sensitive parameter characterizing photon statistics from nonclassical antibunching ($g^{(2)}(0) < 0$) via classical coherent states ($g^{(2)}(0) = 1$) to thermal emissiion $g^{(2)}(0) \approx 2$ \cite{Ana_superradiance_2}. In Fig.~\ref{fig:fig4}(a), we see good quantitative agreement between master equation and the collective mode equations at early times, $t\lesssim 1/\gamma_0$, where it decreases to below one around the peak intensity of photon emission. After the peak in emission, $g^{(2)}$ increases again when the remaining excitation is low and in predominantly stored in subradiant states, which have a small photon emission rate. This creates a small denominator in Eq.~\ref{eq:g2} that results in an increase of $g^{(2)}(0)$. As expected, this feature cannot be very well represented in our truncated model and errors increase significantly at late times.  
For comparison, we also show the two special cases of non-interacting independent emitters ($a\rightarrow \infty$) and Dicke collective superradiance ($a \rightarrow 0$ in the absence of coherent dipole-dipole interactions, $J_\mathrm{nm} = 0$). The non-interacting case leads to an exponential decay of both the photon emission and excited state population, and to a constant $g^{(2)}(0) = 1-1/N$. The Dicke limit, on the other hand, shows the highest possible peak photon emission and super-exponential decay of the excited state population in Fig.~\ref{fig:fig4}(a). The second order correlation can be analytically found at the initial time ($t=0$), $ g^{(2)}(0) = 2-2/N$. As occured for the case of ring geometries, the $g^{(2)}(0)$-function in the Dicke limit decrease at initial times until a minimum is reached. This feature is is discussed below in more detail.

The large deviation of the the $g^{(2)}(0)$ at late times $t \gtrsim \gamma_0/10$ shown by the cumulant expansion in the collective mode basis as opposed to the master equation require further investigation. A first analysis indicates that these growing errors are connected to the relevance of very subradiant states in the dynamics. One possibile improvement consists on the extension of the equations in Eq.~\ref{eq:collective} to include the time evolution of $\langle \hat{S}_{\boldsymbol{0}}^\dagger \hat{S}_{\boldsymbol{0}}^\dagger \hat{S}_{\boldsymbol{0}} \hat{S}_{\boldsymbol{0}} \hat{S}_{\boldsymbol{0}}^{ee}\rangle$. We show in Appendix~\ref{supp:cumulant-collective} that the cumulant expansion of this expectation is smaller than the exact value. This discrepancy could explain in turn the underestimation observed in Fig.~\ref{fig:fig4}(a).
The relative errors of the peak photon emission $\Delta p_\mathrm{out}^\mathrm{max}$ and the second order correlation function $\Delta g^{(2)}(0)$ at peak photon emission both tend to decrease towards as a function of the emitter number N. Note, that in (a)-(c) no modes are truncated, thus any error stems from the cumulant expansion performed in Eq.~\ref{eq:collective}.

Finally, we show $\Delta p_\mathrm{out}^\mathrm{max}$ as a function of the number of truncated collective modes for $N=12$ emitters in Fig.~\ref{fig:fig4}(d). Notably, the relative error remains close to zero if at most $7$ out of the $12$ total modes are trucnated. In other words, the magnitude of the superradiant peak only diverges from the exact value when $8$ or more modes are truncated. This constitutes a $50 \%$ reduction of the number of modes that need to be considered.
\subsection*{Second order correlation function $\boldsymbol{g^{(2)}(\tau=0)}$}
The second order correlation with zero time delay $g^{(2)}(\tau=0)$ in Eq.~\ref{eq:g2} gives a measure of emitting two photons simultaneously from an ensemble of emitters. In other words, it is a measure of how much photons bunch, i.e., how much more likely is it, if we measure one photon, to have a second one arrive at the same time. It has been shown previously~\cite{Ana_superradiance_2} that the value of $g^{(2)}(0)$, at $t=0$ predicts whether the system will exhibit a superradiant outburst of radiation or not. The minimal condition for this to happen is $g^{{(2)}}(0)|_{t=0} > 1$~\cite{Ana_superradiance_2,masson2024dicke,Robicheaux_superradiance}. For subwavelength rings, Fig.~\ref{fig:fig4}(a) shows that the $g^{(2)}(0)$-function is above unity at $t=0$, but then decreases at later times until a local minimum is reached after peak photon emission.

Utilizing the collective equations of motion for Eq.~\ref{eq:collective} and provided in Appendix~\ref{supp:cumulant-collective}, we can evalute the time derivate at $t=0$ and obtain
\begin{align} \label{derivative-g2}
    \frac{d}{dt} g^{(2)}(0) \Big|_{t=0} &= \frac{2}{N^2}\sum_{\boldsymbol{\mu}} \Gamma_{\boldsymbol{\mu}}^3 - \frac{4}{N^2} \sum_{\boldsymbol{\mu}} \Gamma_{\boldsymbol{\mu}}^2 \nonumber \\
    &- \frac{2}{N^3}\Big(\sum_{\boldsymbol{\mu}} \Gamma_{\boldsymbol{\mu}}^2 \Big)^2 +\frac{4}{N} < 0.
\end{align}
The derivative is confined to the interval $[-4+4/N,0)$, thus confirming that the $g^{(2)}(0)$ will always show an initial decrease. In the Dicke limit ($a = 0$)~\cite{Dicke_originalpaper}, where $\Gamma_{\boldsymbol{\mu}}=\delta_{\boldsymbol{0 \mu}} N$, it equals $-4(1-1/N)$. In the opposit limit of independent decay ($a \rightarrow \infty$), $\Gamma_{\boldsymbol{\mu}} \rightarrow \gamma_0$ for all $\boldsymbol{\mu}$ and Eq.~\ref{derivative-g2} approaches zero.
Then, the second order correlation remains constant and equals $1-1/N$ throughout the time evolution, as shown in Fig.~\ref{fig:fig4}(a). In the presence of interactions, however, $g^{(2)}(\tau=0)$ always decreases at $t=0$, and for the cases shown in Fig.~\ref{fig:fig4}(a), will reach a local minimum after peak photon emission. In a physical picture, this means that the simultaneous emission of two photons becomes less and less likely after $t=0$ and until after peak photon emission, even though the photon emission itself keeps increasing. At late times, $g^{(2)}(\tau=0)$ increases due to the rapid decrease of the emitted intensity in the denominator of Eq.~(\ref{eq:g2}). We believe, that this warrants further investigation and discuss possible directions in Section~\ref{sec:conclusions}.
\subsection*{Large scale ordered arrays}
We apply the collective mode truncation to ordered arrays involving large numbers of emitters and provide a comparison with second and third order cumulant expansions in the individual spin basis as a function of $N$. In Fig.~\ref{fig:fig5} we truncate the number of collective modes used to solve Eq.~\ref{eom} to a total of $\tilde{N}$ modes  satisfying $\Gamma_{\boldsymbol{{\mu}}}>\gamma_0/N$.
In Fig.~\ref{fig:fig5}(a) the peak photon emission $p_\mathrm{out}^\mathrm{max}$ for the chain array saturates at $\approx 1.1N$ for the third order cumulant expansion and collective mode truncation. The second order cumulant expansion meanwhile shows an overestimation of approximately $5 \%$. The second order correlation $g^{(2)}(0)$ is calculated with the collective mode truncation and shows convergence to unity at peak photon emission. The number of equations, i.e. numerical complexity for the collective mode truncation is well desribed by $(2Na/\lambda_0)^3$, as previously obtained inFig.~\ref{fig:fig2}(c). As a results, there are approximately two orders of magnitude less equations to be solved when using the collective spin basis with mode truncation than when using the third order cumulant expansion in the individual spin basis for arrays with spacing $a=0.15 \lambda_0$.

The peak photon emission for the two-dimensional square array in Fig.~\ref{fig:fig5}(b) shows again excellent agreement between the third order cumulant expansion and the collective mode truncation. The second order cumulant expansion meanwhile shows now an overestimation of approximately $10$-$20 \%$. The numerical complexity as a function of $N$ for the collective mode truncation is now slowly converging to the analytical value obtained earlier, $(N\pi a^2/\lambda_0^2)^3$. As a consequence, the number of equations that have to be solved as compared to the third order cumulant expansion in the individual spin basis are reduced by a factor $\sim 10^3$ at $a=0.15\lambda_0$ and for large $N$. 

\section{Discussion and outlook} \label{sec:conclusions}
We have presented an efficient numerical technique for simulating large numbers of dipole-coupled quantum emitters in ordered arrays in the presence of collective long-range decay.
For subwavelength array spacings, we have demonstrated the emergence of collective super- and subradiant modes by transforming the individual spin operators into a collective spin basis. Subsequently, we have applied a cumulant expansion to the Heisenberg equations of motion to derive a closed set of equations that approximatelly describes the collective decay dynamics.

Crucially, a majority of collective modes for subwavelength spacings are highly subradiant, and therefore do not participate significantly in the radiant dissipative dynamics. By neglecting or truncating these modes, the numerical complexity of the system of equations can be strongly reduced while ensuring an accurate description of the dissipative time evolution of the quantum many-body system.

We have illustrated and benchmarked this collective mode truncation method for various geometries by comparing the resulting dynamics with that obtained by solving small systems of up to $14$ emitters exactly or by applying a cumulant expansion based on the individual spin basis for larger numbers of emitter. 
The method presented in this work allows to make accurate predictions about excited state populations, photon emission rates and second-order correlations, highly relevant for many quantum many-body experiments and dipolar spin ensemble experiments~\cite{ferioli2022observation,baier2016extended}. We have also studied the photon statistics of the emitted light by analyzing the second order intensity correlation function with zero time delay, $g^{(2)}(\tau=0)$ , thereby characterizing two photon processes. Unexpectedly, in the presence of dipole-dipole interactions, this quantity showed a decrease until after peak photon emission. Future studies should further elucidte the behavior of the second- and higher-order correlation function at $t\gtrsim 0$, as it provides key insights into the internal correlations of the ensemble of emitters~\cite{masson2020many,Ana_superradiance_2}.

Future implementations of the collective mode truncation method might include incoherent driving of the individual emitters~\cite{holzinger2020nanoscale}, emitters with multiple energy levels~\cite{masson2024dicke}, initially coherent states~\cite{masson2020many} and coherent driving~\cite{ferioli2021laser,ferioli2022observation} to study, for instance, the generalization of the driven Dicke model in free space~\cite{ferioli2022observation,Ostermann_DDM}. This method can also be leveraged to simulate the collective dissipative dynamics of emitter arrays coupled to electromagnetic environments different than free space, such as waveguides~\cite{sheremet2023waveguide,pennetta2022collective,tebbenjohanns2024predicting} (see Appendix~\ref{appendix:finite}).

Finally, we would like to point out that the collective basis presented in this work also opens the possibility to explore subradiant dynamics in a perturbative fashion. Here, subradiant modes with a specific quasi-momentum can be selectively added to or removed from the existing system of equations, thereby allowing to study their role in the dissipative dynamics.

\section*{Acknowledgements}
We thank Anna Bychek and Claudiu Genes for useful discussions. R.H. acknowledges funding by the Austrian Science Fund (FWF) 10.55776/W1259. O.R.B. acknowledges support from Fundación Mauricio y Carlota Botton and from Fundació Bancaria “la Caixa” (LCF/BQ/AA18/11680093). SFY thanks the NSF through PHY-2207972, the CUA PFC PHY-2317134, and the Q-SEnSE QLCI OMA-2016244.
\newline
\section*{Data availability}
All data and code in this manuscript is available upon reasonable request.

\bibliography{references}

\begin{thebibliography}{70}%
\makeatletter
\providecommand \@ifxundefined [1]{%
 \@ifx{#1\undefined}
}%
\providecommand \@ifnum [1]{%
 \ifnum #1\expandafter \@firstoftwo
 \else \expandafter \@secondoftwo
 \fi
}%
\providecommand \@ifx [1]{%
 \ifx #1\expandafter \@firstoftwo
 \else \expandafter \@secondoftwo
 \fi
}%
\providecommand \natexlab [1]{#1}%
\providecommand \enquote  [1]{``#1''}%
\providecommand \bibnamefont  [1]{#1}%
\providecommand \bibfnamefont [1]{#1}%
\providecommand \citenamefont [1]{#1}%
\providecommand \href@noop [0]{\@secondoftwo}%
\providecommand \href [0]{\begingroup \@sanitize@url \@href}%
\providecommand \@href[1]{\@@startlink{#1}\@@href}%
\providecommand \@@href[1]{\endgroup#1\@@endlink}%
\providecommand \@sanitize@url [0]{\catcode `\\12\catcode `\$12\catcode `\&12\catcode `\#12\catcode `\^12\catcode `\_12\catcode `\%12\relax}%
\providecommand \@@startlink[1]{}%
\providecommand \@@endlink[0]{}%
\providecommand \url  [0]{\begingroup\@sanitize@url \@url }%
\providecommand \@url [1]{\endgroup\@href {#1}{\urlprefix }}%
\providecommand \urlprefix  [0]{URL }%
\providecommand \Eprint [0]{\href }%
\providecommand \doibase [0]{https://doi.org/}%
\providecommand \selectlanguage [0]{\@gobble}%
\providecommand \bibinfo  [0]{\@secondoftwo}%
\providecommand \bibfield  [0]{\@secondoftwo}%
\providecommand \translation [1]{[#1]}%
\providecommand \BibitemOpen [0]{}%
\providecommand \bibitemStop [0]{}%
\providecommand \bibitemNoStop [0]{.\EOS\space}%
\providecommand \EOS [0]{\spacefactor3000\relax}%
\providecommand \BibitemShut  [1]{\csname bibitem#1\endcsname}%
\let\auto@bib@innerbib\@empty
\bibitem [{\citenamefont {Reitz}\ \emph {et~al.}(2022)\citenamefont {Reitz}, \citenamefont {Sommer},\ and\ \citenamefont {Genes}}]{reitz2022cooperative}%
  \BibitemOpen
  \bibfield  {author} {\bibinfo {author} {\bibfnamefont {M.}~\bibnamefont {Reitz}}, \bibinfo {author} {\bibfnamefont {C.}~\bibnamefont {Sommer}},\ and\ \bibinfo {author} {\bibfnamefont {C.}~\bibnamefont {Genes}},\ }\bibfield  {title} {\bibinfo {title} {\href{https://link.aps.org/doi/10.1103/PRXQuantum.3.010201}{Cooperative quantum phenomena in light-matter platforms}},\ }\href@noop {} {\bibfield  {journal} {\bibinfo  {journal} {PRX Quantum}\ }\textbf {\bibinfo {volume} {3}},\ \bibinfo {pages} {010201} (\bibinfo {year} {2022})}\BibitemShut {NoStop}%
\bibitem [{\citenamefont {Masson}\ \emph {et~al.}(2020)\citenamefont {Masson}, \citenamefont {Ferrier-Barbut}, \citenamefont {Orozco}, \citenamefont {Browaeys},\ and\ \citenamefont {Asenjo-Garcia}}]{masson2020many}%
  \BibitemOpen
  \bibfield  {author} {\bibinfo {author} {\bibfnamefont {S.~J.}\ \bibnamefont {Masson}}, \bibinfo {author} {\bibfnamefont {I.}~\bibnamefont {Ferrier-Barbut}}, \bibinfo {author} {\bibfnamefont {L.~A.}\ \bibnamefont {Orozco}}, \bibinfo {author} {\bibfnamefont {A.}~\bibnamefont {Browaeys}},\ and\ \bibinfo {author} {\bibfnamefont {A.}~\bibnamefont {Asenjo-Garcia}},\ }\bibfield  {title} {\bibinfo {title} {\href{https://journals.aps.org/prl/abstract/10.1103/PhysRevLett.125.263601}{Many-body signatures of collective decay in atomic chains}},\ }\href@noop {} {\bibfield  {journal} {\bibinfo  {journal} {Physical review letters}\ }\textbf {\bibinfo {volume} {125}},\ \bibinfo {pages} {263601} (\bibinfo {year} {2020})}\BibitemShut {NoStop}%
\bibitem [{\citenamefont {Masson}\ and\ \citenamefont {Asenjo-Garcia}(2022)}]{Ana_superradiance_2}%
  \BibitemOpen
  \bibfield  {author} {\bibinfo {author} {\bibfnamefont {S.~J.}\ \bibnamefont {Masson}}\ and\ \bibinfo {author} {\bibfnamefont {A.}~\bibnamefont {Asenjo-Garcia}},\ }\bibfield  {title} {\bibinfo {title} {Universality of dicke superradiance in arrays of quantum emitters},\ }\href {https://doi.org/10.1038/s41467-022-29805-4} {\bibfield  {journal} {\bibinfo  {journal} {Nature Communications}\ }\textbf {\bibinfo {volume} {13}},\ \bibinfo {pages} {2285} (\bibinfo {year} {2022})}\BibitemShut {NoStop}%
\bibitem [{\citenamefont {Sutherland}\ and\ \citenamefont {Robicheaux}(2017)}]{sutherland2017superradiance}%
  \BibitemOpen
  \bibfield  {author} {\bibinfo {author} {\bibfnamefont {R.~T.}\ \bibnamefont {Sutherland}}\ and\ \bibinfo {author} {\bibfnamefont {F.}~\bibnamefont {Robicheaux}},\ }\bibfield  {title} {\bibinfo {title} {\href{https://journals.aps.org/pra/abstract/10.1103/PhysRevA.95.033839}{Superradiance in inverted multilevel atomic clouds}},\ }\href@noop {} {\bibfield  {journal} {\bibinfo  {journal} {Physical Review A}\ }\textbf {\bibinfo {volume} {95}},\ \bibinfo {pages} {033839} (\bibinfo {year} {2017})}\BibitemShut {NoStop}%
\bibitem [{\citenamefont {Rubies-Bigorda}\ and\ \citenamefont {Yelin}(2022)}]{rubies2022superradiance}%
  \BibitemOpen
  \bibfield  {author} {\bibinfo {author} {\bibfnamefont {O.}~\bibnamefont {Rubies-Bigorda}}\ and\ \bibinfo {author} {\bibfnamefont {S.~F.}\ \bibnamefont {Yelin}},\ }\bibfield  {title} {\bibinfo {title} {Superradiance and subradiance in inverted atomic arrays},\ }\href {https://doi.org/10.1103/PhysRevA.106.053717} {\bibfield  {journal} {\bibinfo  {journal} {Phys. Rev. A}\ }\textbf {\bibinfo {volume} {106}},\ \bibinfo {pages} {053717} (\bibinfo {year} {2022})}\BibitemShut {NoStop}%
\bibitem [{\citenamefont {Parmee}\ and\ \citenamefont {Ruostekoski}(2020)}]{parmee2020signatures}%
  \BibitemOpen
  \bibfield  {author} {\bibinfo {author} {\bibfnamefont {C.~D.}\ \bibnamefont {Parmee}}\ and\ \bibinfo {author} {\bibfnamefont {J.}~\bibnamefont {Ruostekoski}},\ }\bibfield  {title} {\bibinfo {title} {\href{https://www.nature.com/articles/s42005-020-00476-1}{Signatures of optical phase transitions in superradiant and subradiant atomic arrays}},\ }\href@noop {} {\bibfield  {journal} {\bibinfo  {journal} {Communications Physics}\ }\textbf {\bibinfo {volume} {3}},\ \bibinfo {pages} {205} (\bibinfo {year} {2020})}\BibitemShut {NoStop}%
\bibitem [{\citenamefont {Holzinger}\ \emph {et~al.}(2021)\citenamefont {Holzinger}, \citenamefont {Moreno-Cardoner},\ and\ \citenamefont {Ritsch}}]{holzinger2021nanoscale}%
  \BibitemOpen
  \bibfield  {author} {\bibinfo {author} {\bibfnamefont {R.}~\bibnamefont {Holzinger}}, \bibinfo {author} {\bibfnamefont {M.}~\bibnamefont {Moreno-Cardoner}},\ and\ \bibinfo {author} {\bibfnamefont {H.}~\bibnamefont {Ritsch}},\ }\bibfield  {title} {\bibinfo {title} {\href{https://pubs.aip.org/aip/apl/article-abstract/119/2/024002/1023161/Nanoscale-continuous-quantum-light-sources-based?redirectedFrom=fulltext}{Nanoscale continuous quantum light sources based on driven dipole emitter arrays}},\ }\href@noop {} {\bibfield  {journal} {\bibinfo  {journal} {Applied Physics Letters}\ }\textbf {\bibinfo {volume} {119}} (\bibinfo {year} {2021})}\BibitemShut {NoStop}%
\bibitem [{\citenamefont {Ferioli}\ \emph {et~al.}(2023)\citenamefont {Ferioli}, \citenamefont {Glicenstein}, \citenamefont {Ferrier-Barbut},\ and\ \citenamefont {Browaeys}}]{ferioli2022observation}%
  \BibitemOpen
  \bibfield  {author} {\bibinfo {author} {\bibfnamefont {G.}~\bibnamefont {Ferioli}}, \bibinfo {author} {\bibfnamefont {A.}~\bibnamefont {Glicenstein}}, \bibinfo {author} {\bibfnamefont {I.}~\bibnamefont {Ferrier-Barbut}},\ and\ \bibinfo {author} {\bibfnamefont {A.}~\bibnamefont {Browaeys}},\ }\bibfield  {title} {\bibinfo {title} {A non-equilibrium superradiant phase transition in free space},\ }\href {https://doi.org/10.1038/s41567-023-02064-w} {\bibfield  {journal} {\bibinfo  {journal} {Nature Physics}\ }\textbf {\bibinfo {volume} {19}},\ \bibinfo {pages} {1345} (\bibinfo {year} {2023})}\BibitemShut {NoStop}%
\bibitem [{\citenamefont {Glicenstein}\ \emph {et~al.}(2022)\citenamefont {Glicenstein}, \citenamefont {Ferioli}, \citenamefont {Browaeys},\ and\ \citenamefont {Ferrier-Barbut}}]{glicenstein2022superradiance}%
  \BibitemOpen
  \bibfield  {author} {\bibinfo {author} {\bibfnamefont {A.}~\bibnamefont {Glicenstein}}, \bibinfo {author} {\bibfnamefont {G.}~\bibnamefont {Ferioli}}, \bibinfo {author} {\bibfnamefont {A.}~\bibnamefont {Browaeys}},\ and\ \bibinfo {author} {\bibfnamefont {I.}~\bibnamefont {Ferrier-Barbut}},\ }\bibfield  {title} {\bibinfo {title} {\href{https://opg.optica.org/ol/abstract.cfm?uri=ol-47-6-1541}{From superradiance to subradiance: exploring the many-body Dicke ladder}},\ }\href@noop {} {\bibfield  {journal} {\bibinfo  {journal} {Optics Letters}\ }\textbf {\bibinfo {volume} {47}},\ \bibinfo {pages} {1541} (\bibinfo {year} {2022})}\BibitemShut {NoStop}%
\bibitem [{\citenamefont {Ferioli}\ \emph {et~al.}(2021{\natexlab{a}})\citenamefont {Ferioli}, \citenamefont {Glicenstein}, \citenamefont {Robicheaux}, \citenamefont {Sutherland}, \citenamefont {Browaeys},\ and\ \citenamefont {Ferrier-Barbut}}]{ferioli2021laser}%
  \BibitemOpen
  \bibfield  {author} {\bibinfo {author} {\bibfnamefont {G.}~\bibnamefont {Ferioli}}, \bibinfo {author} {\bibfnamefont {A.}~\bibnamefont {Glicenstein}}, \bibinfo {author} {\bibfnamefont {F.}~\bibnamefont {Robicheaux}}, \bibinfo {author} {\bibfnamefont {R.}~\bibnamefont {Sutherland}}, \bibinfo {author} {\bibfnamefont {A.}~\bibnamefont {Browaeys}},\ and\ \bibinfo {author} {\bibfnamefont {I.}~\bibnamefont {Ferrier-Barbut}},\ }\bibfield  {title} {\bibinfo {title} {\href{https://journals.aps.org/prl/abstract/10.1103/PhysRevLett.127.243602}{Laser-driven superradiant ensembles of two-level atoms near Dicke regime}},\ }\href@noop {} {\bibfield  {journal} {\bibinfo  {journal} {Physical Review Letters}\ }\textbf {\bibinfo {volume} {127}},\ \bibinfo {pages} {243602} (\bibinfo {year} {2021}{\natexlab{a}})}\BibitemShut {NoStop}%
\bibitem [{\citenamefont {Ferioli}\ \emph {et~al.}(2021{\natexlab{b}})\citenamefont {Ferioli}, \citenamefont {Glicenstein}, \citenamefont {Henriet}, \citenamefont {Ferrier-Barbut},\ and\ \citenamefont {Browaeys}}]{ferioli2021storage}%
  \BibitemOpen
  \bibfield  {author} {\bibinfo {author} {\bibfnamefont {G.}~\bibnamefont {Ferioli}}, \bibinfo {author} {\bibfnamefont {A.}~\bibnamefont {Glicenstein}}, \bibinfo {author} {\bibfnamefont {L.}~\bibnamefont {Henriet}}, \bibinfo {author} {\bibfnamefont {I.}~\bibnamefont {Ferrier-Barbut}},\ and\ \bibinfo {author} {\bibfnamefont {A.}~\bibnamefont {Browaeys}},\ }\bibfield  {title} {\bibinfo {title} {\href{https://journals.aps.org/prx/abstract/10.1103/PhysRevX.11.021031}{Storage and release of subradiant excitations in a dense atomic cloud}},\ }\href@noop {} {\bibfield  {journal} {\bibinfo  {journal} {Physical Review X}\ }\textbf {\bibinfo {volume} {11}},\ \bibinfo {pages} {021031} (\bibinfo {year} {2021}{\natexlab{b}})}\BibitemShut {NoStop}%
\bibitem [{\citenamefont {Inouye}\ \emph {et~al.}(1999)\citenamefont {Inouye}, \citenamefont {Chikkatur}, \citenamefont {Stamper-Kurn}, \citenamefont {Stenger}, \citenamefont {Pritchard},\ and\ \citenamefont {Ketterle}}]{inouye1999superradiant}%
  \BibitemOpen
  \bibfield  {author} {\bibinfo {author} {\bibfnamefont {S.}~\bibnamefont {Inouye}}, \bibinfo {author} {\bibfnamefont {A.}~\bibnamefont {Chikkatur}}, \bibinfo {author} {\bibfnamefont {D.}~\bibnamefont {Stamper-Kurn}}, \bibinfo {author} {\bibfnamefont {J.}~\bibnamefont {Stenger}}, \bibinfo {author} {\bibfnamefont {D.}~\bibnamefont {Pritchard}},\ and\ \bibinfo {author} {\bibfnamefont {W.}~\bibnamefont {Ketterle}},\ }\bibfield  {title} {\bibinfo {title} {\href{https://www.science.org/doi/abs/10.1126/science.285.5427.571}{Superradiant Rayleigh scattering from a Bose-Einstein condensate}},\ }\href@noop {} {\bibfield  {journal} {\bibinfo  {journal} {Science}\ }\textbf {\bibinfo {volume} {285}},\ \bibinfo {pages} {571} (\bibinfo {year} {1999})}\BibitemShut {NoStop}%
\bibitem [{\citenamefont {Grimes}\ \emph {et~al.}(2017)\citenamefont {Grimes}, \citenamefont {Coy}, \citenamefont {Barnum}, \citenamefont {Zhou}, \citenamefont {Yelin},\ and\ \citenamefont {Field}}]{grimes2017direct}%
  \BibitemOpen
  \bibfield  {author} {\bibinfo {author} {\bibfnamefont {D.~D.}\ \bibnamefont {Grimes}}, \bibinfo {author} {\bibfnamefont {S.~L.}\ \bibnamefont {Coy}}, \bibinfo {author} {\bibfnamefont {T.~J.}\ \bibnamefont {Barnum}}, \bibinfo {author} {\bibfnamefont {Y.}~\bibnamefont {Zhou}}, \bibinfo {author} {\bibfnamefont {S.~F.}\ \bibnamefont {Yelin}},\ and\ \bibinfo {author} {\bibfnamefont {R.~W.}\ \bibnamefont {Field}},\ }\bibfield  {title} {\bibinfo {title} {\href{https://journals.aps.org/pra/abstract/10.1103/PhysRevA.95.043818}{Direct single-shot observation of millimeter-wave superradiance in Rydberg-Rydberg transitions}},\ }\href {https://doi.org/10.1103/PhysRevA.95.043818} {\bibfield  {journal} {\bibinfo  {journal} {Phys. Rev. A}\ }\textbf {\bibinfo {volume} {95}},\ \bibinfo {pages} {043818} (\bibinfo {year} {2017})}\BibitemShut {NoStop}%
\bibitem [{\citenamefont {Kaluzny}\ \emph {et~al.}(1983)\citenamefont {Kaluzny}, \citenamefont {Goy}, \citenamefont {Gross}, \citenamefont {Raimond},\ and\ \citenamefont {Haroche}}]{kaluzny1983observation}%
  \BibitemOpen
  \bibfield  {author} {\bibinfo {author} {\bibfnamefont {Y.}~\bibnamefont {Kaluzny}}, \bibinfo {author} {\bibfnamefont {P.}~\bibnamefont {Goy}}, \bibinfo {author} {\bibfnamefont {M.}~\bibnamefont {Gross}}, \bibinfo {author} {\bibfnamefont {J.~M.}\ \bibnamefont {Raimond}},\ and\ \bibinfo {author} {\bibfnamefont {S.}~\bibnamefont {Haroche}},\ }\bibfield  {title} {\bibinfo {title} {\href{https://journals.aps.org/prl/abstract/10.1103/PhysRevLett.51.1175}{Observation of Self-Induced Rabi Oscillations in Two-Level Atoms Excited Inside a Resonant Cavity: The Ringing Regime of Superradiance}},\ }\href {https://doi.org/10.1103/PhysRevLett.51.1175} {\bibfield  {journal} {\bibinfo  {journal} {Phys. Rev. Lett.}\ }\textbf {\bibinfo {volume} {51}},\ \bibinfo {pages} {1175} (\bibinfo {year} {1983})}\BibitemShut {NoStop}%
\bibitem [{\citenamefont {Chen}\ \emph {et~al.}(2018)\citenamefont {Chen}, \citenamefont {Wang}, \citenamefont {Meng}, \citenamefont {Huang}, \citenamefont {Cai}, \citenamefont {Wang}, \citenamefont {Zhu},\ and\ \citenamefont {Zhang}}]{chen2018experimental}%
  \BibitemOpen
  \bibfield  {author} {\bibinfo {author} {\bibfnamefont {L.}~\bibnamefont {Chen}}, \bibinfo {author} {\bibfnamefont {P.}~\bibnamefont {Wang}}, \bibinfo {author} {\bibfnamefont {Z.}~\bibnamefont {Meng}}, \bibinfo {author} {\bibfnamefont {L.}~\bibnamefont {Huang}}, \bibinfo {author} {\bibfnamefont {H.}~\bibnamefont {Cai}}, \bibinfo {author} {\bibfnamefont {D.-W.}\ \bibnamefont {Wang}}, \bibinfo {author} {\bibfnamefont {S.-Y.}\ \bibnamefont {Zhu}},\ and\ \bibinfo {author} {\bibfnamefont {J.}~\bibnamefont {Zhang}},\ }\bibfield  {title} {\bibinfo {title} {\href{https://journals.aps.org/prl/abstract/10.1103/PhysRevLett.120.193601}{Experimental Observation of One-Dimensional Superradiance Lattices in Ultracold Atoms}},\ }\href {https://doi.org/10.1103/PhysRevLett.120.193601} {\bibfield  {journal} {\bibinfo  {journal} {Phys. Rev. Lett.}\ }\textbf {\bibinfo {volume} {120}},\ \bibinfo {pages} {193601} (\bibinfo {year} {2018})}\BibitemShut {NoStop}%
\bibitem [{\citenamefont {Pellegrino}\ \emph {et~al.}(2014)\citenamefont {Pellegrino}, \citenamefont {Bourgain}, \citenamefont {Jennewein}, \citenamefont {Sortais}, \citenamefont {Browaeys}, \citenamefont {Jenkins},\ and\ \citenamefont {Ruostekoski}}]{pellegrino2014observation}%
  \BibitemOpen
  \bibfield  {author} {\bibinfo {author} {\bibfnamefont {J.}~\bibnamefont {Pellegrino}}, \bibinfo {author} {\bibfnamefont {R.}~\bibnamefont {Bourgain}}, \bibinfo {author} {\bibfnamefont {S.}~\bibnamefont {Jennewein}}, \bibinfo {author} {\bibfnamefont {Y.~R.}\ \bibnamefont {Sortais}}, \bibinfo {author} {\bibfnamefont {A.}~\bibnamefont {Browaeys}}, \bibinfo {author} {\bibfnamefont {S.}~\bibnamefont {Jenkins}},\ and\ \bibinfo {author} {\bibfnamefont {J.}~\bibnamefont {Ruostekoski}},\ }\bibfield  {title} {\bibinfo {title} {\href{https://journals.aps.org/prl/abstract/10.1103/PhysRevLett.113.133602}{Observation of suppression of light scattering induced by dipole-dipole interactions in a cold-atom ensemble}},\ }\href@noop {} {\bibfield  {journal} {\bibinfo  {journal} {Physical review letters}\ }\textbf {\bibinfo {volume} {113}},\ \bibinfo {pages} {133602} (\bibinfo {year} {2014})}\BibitemShut {NoStop}%
\bibitem [{\citenamefont {Chang}\ \emph {et~al.}(2018)\citenamefont {Chang}, \citenamefont {Douglas}, \citenamefont {Gonz\'alez-Tudela}, \citenamefont {Hung},\ and\ \citenamefont {Kimble}}]{Chang2018}%
  \BibitemOpen
  \bibfield  {author} {\bibinfo {author} {\bibfnamefont {D.~E.}\ \bibnamefont {Chang}}, \bibinfo {author} {\bibfnamefont {J.~S.}\ \bibnamefont {Douglas}}, \bibinfo {author} {\bibfnamefont {A.}~\bibnamefont {Gonz\'alez-Tudela}}, \bibinfo {author} {\bibfnamefont {C.-L.}\ \bibnamefont {Hung}},\ and\ \bibinfo {author} {\bibfnamefont {H.~J.}\ \bibnamefont {Kimble}},\ }\bibfield  {title} {\bibinfo {title} {Colloquium: Quantum matter built from nanoscopic lattices of atoms and photons},\ }\href {https://doi.org/10.1103/RevModPhys.90.031002} {\bibfield  {journal} {\bibinfo  {journal} {Rev. Mod. Phys.}\ }\textbf {\bibinfo {volume} {90}},\ \bibinfo {pages} {031002} (\bibinfo {year} {2018})}\BibitemShut {NoStop}%
\bibitem [{\citenamefont {Huang}\ \emph {et~al.}(2023)\citenamefont {Huang}, \citenamefont {Yuan}, \citenamefont {Holman}, \citenamefont {Kwon}, \citenamefont {Masson}, \citenamefont {Gutierrez-Jauregui}, \citenamefont {Asenjo-Garcia}, \citenamefont {Will},\ and\ \citenamefont {Yu}}]{HUANG2023100470}%
  \BibitemOpen
  \bibfield  {author} {\bibinfo {author} {\bibfnamefont {X.}~\bibnamefont {Huang}}, \bibinfo {author} {\bibfnamefont {W.}~\bibnamefont {Yuan}}, \bibinfo {author} {\bibfnamefont {A.}~\bibnamefont {Holman}}, \bibinfo {author} {\bibfnamefont {M.}~\bibnamefont {Kwon}}, \bibinfo {author} {\bibfnamefont {S.~J.}\ \bibnamefont {Masson}}, \bibinfo {author} {\bibfnamefont {R.}~\bibnamefont {Gutierrez-Jauregui}}, \bibinfo {author} {\bibfnamefont {A.}~\bibnamefont {Asenjo-Garcia}}, \bibinfo {author} {\bibfnamefont {S.}~\bibnamefont {Will}},\ and\ \bibinfo {author} {\bibfnamefont {N.}~\bibnamefont {Yu}},\ }\bibfield  {title} {\bibinfo {title} {Metasurface holographic optical traps for ultracold atoms},\ }\href {https://doi.org/https://doi.org/10.1016/j.pquantelec.2023.100470} {\bibfield  {journal} {\bibinfo  {journal} {Progress in Quantum Electronics}\ }\textbf {\bibinfo {volume} {89}},\ \bibinfo {pages} {100470} (\bibinfo {year} {2023})}\BibitemShut {NoStop}%
\bibitem [{\citenamefont {Holzinger}\ \emph {et~al.}(2020)\citenamefont {Holzinger}, \citenamefont {Plankensteiner}, \citenamefont {Ostermann},\ and\ \citenamefont {Ritsch}}]{holzinger2020nanoscale}%
  \BibitemOpen
  \bibfield  {author} {\bibinfo {author} {\bibfnamefont {R.}~\bibnamefont {Holzinger}}, \bibinfo {author} {\bibfnamefont {D.}~\bibnamefont {Plankensteiner}}, \bibinfo {author} {\bibfnamefont {L.}~\bibnamefont {Ostermann}},\ and\ \bibinfo {author} {\bibfnamefont {H.}~\bibnamefont {Ritsch}},\ }\bibfield  {title} {\bibinfo {title} {\href{https://journals.aps.org/prl/abstract/10.1103/PhysRevLett.124.253603}{Nanoscale coherent light source}},\ }\href@noop {} {\bibfield  {journal} {\bibinfo  {journal} {Physical Review Letters}\ }\textbf {\bibinfo {volume} {124}},\ \bibinfo {pages} {253603} (\bibinfo {year} {2020})}\BibitemShut {NoStop}%
\bibitem [{\citenamefont {Cox}\ \emph {et~al.}(2016)\citenamefont {Cox}, \citenamefont {Greve}, \citenamefont {Weiner},\ and\ \citenamefont {Thompson}}]{cox2016deterministic}%
  \BibitemOpen
  \bibfield  {author} {\bibinfo {author} {\bibfnamefont {K.~C.}\ \bibnamefont {Cox}}, \bibinfo {author} {\bibfnamefont {G.~P.}\ \bibnamefont {Greve}}, \bibinfo {author} {\bibfnamefont {J.~M.}\ \bibnamefont {Weiner}},\ and\ \bibinfo {author} {\bibfnamefont {J.~K.}\ \bibnamefont {Thompson}},\ }\bibfield  {title} {\bibinfo {title} {\href{https://journals.aps.org/prl/abstract/10.1103/PhysRevLett.116.093602}{Deterministic squeezed states with collective measurements and feedback}},\ }\href@noop {} {\bibfield  {journal} {\bibinfo  {journal} {Physical review letters}\ }\textbf {\bibinfo {volume} {116}},\ \bibinfo {pages} {093602} (\bibinfo {year} {2016})}\BibitemShut {NoStop}%
\bibitem [{\citenamefont {Cattaneo}\ \emph {et~al.}(2023)\citenamefont {Cattaneo}, \citenamefont {Rossi}, \citenamefont {Garc\'{\i}a-P\'erez}, \citenamefont {Zambrini},\ and\ \citenamefont {Maniscalco}}]{cattaneo2023quantum}%
  \BibitemOpen
  \bibfield  {author} {\bibinfo {author} {\bibfnamefont {M.}~\bibnamefont {Cattaneo}}, \bibinfo {author} {\bibfnamefont {M.~A.}\ \bibnamefont {Rossi}}, \bibinfo {author} {\bibfnamefont {G.}~\bibnamefont {Garc\'{\i}a-P\'erez}}, \bibinfo {author} {\bibfnamefont {R.}~\bibnamefont {Zambrini}},\ and\ \bibinfo {author} {\bibfnamefont {S.}~\bibnamefont {Maniscalco}},\ }\bibfield  {title} {\bibinfo {title} {Quantum simulation of dissipative collective effects on noisy quantum computers},\ }\href {https://doi.org/10.1103/PRXQuantum.4.010324} {\bibfield  {journal} {\bibinfo  {journal} {PRX Quantum}\ }\textbf {\bibinfo {volume} {4}},\ \bibinfo {pages} {010324} (\bibinfo {year} {2023})}\BibitemShut {NoStop}%
\bibitem [{\citenamefont {Lemberger}\ and\ \citenamefont {Yavuz}(2017)}]{lemberger2017effect}%
  \BibitemOpen
  \bibfield  {author} {\bibinfo {author} {\bibfnamefont {B.}~\bibnamefont {Lemberger}}\ and\ \bibinfo {author} {\bibfnamefont {D.}~\bibnamefont {Yavuz}},\ }\bibfield  {title} {\bibinfo {title} {\href{https://journals.aps.org/pra/abstract/10.1103/PhysRevA.96.062337}{Effect of correlated decay on fault-tolerant quantum computation}},\ }\href@noop {} {\bibfield  {journal} {\bibinfo  {journal} {Physical Review A}\ }\textbf {\bibinfo {volume} {96}},\ \bibinfo {pages} {062337} (\bibinfo {year} {2017})}\BibitemShut {NoStop}%
\bibitem [{\citenamefont {Aharonov}\ \emph {et~al.}(2006)\citenamefont {Aharonov}, \citenamefont {Kitaev},\ and\ \citenamefont {Preskill}}]{aharonov2006fault}%
  \BibitemOpen
  \bibfield  {author} {\bibinfo {author} {\bibfnamefont {D.}~\bibnamefont {Aharonov}}, \bibinfo {author} {\bibfnamefont {A.}~\bibnamefont {Kitaev}},\ and\ \bibinfo {author} {\bibfnamefont {J.}~\bibnamefont {Preskill}},\ }\bibfield  {title} {\bibinfo {title} {\href{https://journals.aps.org/prl/abstract/10.1103/PhysRevLett.96.050504}{Fault-tolerant quantum computation with long-range correlated noise}},\ }\href@noop {} {\bibfield  {journal} {\bibinfo  {journal} {Physical review letters}\ }\textbf {\bibinfo {volume} {96}},\ \bibinfo {pages} {050504} (\bibinfo {year} {2006})}\BibitemShut {NoStop}%
\bibitem [{\citenamefont {Gross}\ and\ \citenamefont {Haroche}(1982)}]{gross_haroche}%
  \BibitemOpen
  \bibfield  {author} {\bibinfo {author} {\bibfnamefont {M.}~\bibnamefont {Gross}}\ and\ \bibinfo {author} {\bibfnamefont {S.}~\bibnamefont {Haroche}},\ }\bibfield  {title} {\bibinfo {title} {Superradiance: An essay on the theory of collective spontaneous emission},\ }\href {https://doi.org/https://doi.org/10.1016/0370-1573(82)90102-8} {\bibfield  {journal} {\bibinfo  {journal} {Physics Reports}\ }\textbf {\bibinfo {volume} {93}},\ \bibinfo {pages} {301} (\bibinfo {year} {1982})}\BibitemShut {NoStop}%
\bibitem [{\citenamefont {Gold}\ \emph {et~al.}(2022)\citenamefont {Gold}, \citenamefont {Huft}, \citenamefont {Young}, \citenamefont {Safari}, \citenamefont {Walker}, \citenamefont {Saffman},\ and\ \citenamefont {Yavuz}}]{gold2022spatial}%
  \BibitemOpen
  \bibfield  {author} {\bibinfo {author} {\bibfnamefont {D.}~\bibnamefont {Gold}}, \bibinfo {author} {\bibfnamefont {P.}~\bibnamefont {Huft}}, \bibinfo {author} {\bibfnamefont {C.}~\bibnamefont {Young}}, \bibinfo {author} {\bibfnamefont {A.}~\bibnamefont {Safari}}, \bibinfo {author} {\bibfnamefont {T.}~\bibnamefont {Walker}}, \bibinfo {author} {\bibfnamefont {M.}~\bibnamefont {Saffman}},\ and\ \bibinfo {author} {\bibfnamefont {D.}~\bibnamefont {Yavuz}},\ }\bibfield  {title} {\bibinfo {title} {\href{https://link.aps.org/doi/10.1103/PRXQuantum.3.010338}{Spatial coherence of light in collective spontaneous emission}},\ }\href@noop {} {\bibfield  {journal} {\bibinfo  {journal} {PRX Quantum}\ }\textbf {\bibinfo {volume} {3}},\ \bibinfo {pages} {010338} (\bibinfo {year} {2022})}\BibitemShut {NoStop}%
\bibitem [{\citenamefont {Gyger}\ \emph {et~al.}(2024)\citenamefont {Gyger}, \citenamefont {Ammenwerth}, \citenamefont {Tao}, \citenamefont {Timme}, \citenamefont {Snigirev}, \citenamefont {Bloch},\ and\ \citenamefont {Zeiher}}]{gyger2024continuous}%
  \BibitemOpen
  \bibfield  {author} {\bibinfo {author} {\bibfnamefont {F.}~\bibnamefont {Gyger}}, \bibinfo {author} {\bibfnamefont {M.}~\bibnamefont {Ammenwerth}}, \bibinfo {author} {\bibfnamefont {R.}~\bibnamefont {Tao}}, \bibinfo {author} {\bibfnamefont {H.}~\bibnamefont {Timme}}, \bibinfo {author} {\bibfnamefont {S.}~\bibnamefont {Snigirev}}, \bibinfo {author} {\bibfnamefont {I.}~\bibnamefont {Bloch}},\ and\ \bibinfo {author} {\bibfnamefont {J.}~\bibnamefont {Zeiher}},\ }\bibfield  {title} {\bibinfo {title} {\href{https://journals.aps.org/prresearch/abstract/10.1103/PhysRevResearch.6.033104}{Continuous operation of large-scale atom arrays in optical lattices}},\ }\href@noop {} {\bibfield  {journal} {\bibinfo  {journal} {Physical Review Research}\ }\textbf {\bibinfo {volume} {6}},\ \bibinfo {pages} {033104} (\bibinfo {year} {2024})}\BibitemShut {NoStop}%
\bibitem [{\citenamefont {Norcia}\ \emph {et~al.}(2024)\citenamefont {Norcia}, \citenamefont {Kim}, \citenamefont {Cairncross}, \citenamefont {Stone}, \citenamefont {Ryou}, \citenamefont {Jaffe}, \citenamefont {Brown}, \citenamefont {Barnes}, \citenamefont {Battaglino}, \citenamefont {Bohdanowicz} \emph {et~al.}}]{norcia2024iterative}%
  \BibitemOpen
  \bibfield  {author} {\bibinfo {author} {\bibfnamefont {M.}~\bibnamefont {Norcia}}, \bibinfo {author} {\bibfnamefont {H.}~\bibnamefont {Kim}}, \bibinfo {author} {\bibfnamefont {W.}~\bibnamefont {Cairncross}}, \bibinfo {author} {\bibfnamefont {M.}~\bibnamefont {Stone}}, \bibinfo {author} {\bibfnamefont {A.}~\bibnamefont {Ryou}}, \bibinfo {author} {\bibfnamefont {M.}~\bibnamefont {Jaffe}}, \bibinfo {author} {\bibfnamefont {M.}~\bibnamefont {Brown}}, \bibinfo {author} {\bibfnamefont {K.}~\bibnamefont {Barnes}}, \bibinfo {author} {\bibfnamefont {P.}~\bibnamefont {Battaglino}}, \bibinfo {author} {\bibfnamefont {T.}~\bibnamefont {Bohdanowicz}}, \emph {et~al.},\ }\bibfield  {title} {\bibinfo {title} {\href{https://link.aps.org/doi/10.1103/PRXQuantum.5.030316}{Iterative Assembly of 171 Yb Atom Arrays with Cavity-Enhanced Optical Lattices}},\ }\href@noop {} {\bibfield  {journal} {\bibinfo  {journal} {PRX Quantum}\ }\textbf {\bibinfo {volume} {5}},\ \bibinfo {pages} {030316} (\bibinfo {year} {2024})}\BibitemShut
  {NoStop}%
\bibitem [{\citenamefont {Patscheider}\ \emph {et~al.}(2020)\citenamefont {Patscheider}, \citenamefont {Zhu}, \citenamefont {Chomaz}, \citenamefont {Petter}, \citenamefont {Baier}, \citenamefont {Rey}, \citenamefont {Ferlaino},\ and\ \citenamefont {Mark}}]{patscheider2020controlling}%
  \BibitemOpen
  \bibfield  {author} {\bibinfo {author} {\bibfnamefont {A.}~\bibnamefont {Patscheider}}, \bibinfo {author} {\bibfnamefont {B.}~\bibnamefont {Zhu}}, \bibinfo {author} {\bibfnamefont {L.}~\bibnamefont {Chomaz}}, \bibinfo {author} {\bibfnamefont {D.}~\bibnamefont {Petter}}, \bibinfo {author} {\bibfnamefont {S.}~\bibnamefont {Baier}}, \bibinfo {author} {\bibfnamefont {A.-M.}\ \bibnamefont {Rey}}, \bibinfo {author} {\bibfnamefont {F.}~\bibnamefont {Ferlaino}},\ and\ \bibinfo {author} {\bibfnamefont {M.}~\bibnamefont {Mark}},\ }\bibfield  {title} {\bibinfo {title} {\href{https://journals.aps.org/prresearch/abstract/10.1103/PhysRevResearch.2.023050}{Controlling dipolar exchange interactions in a dense three-dimensional array of large-spin fermions}},\ }\href@noop {} {\bibfield  {journal} {\bibinfo  {journal} {Physical Review Research}\ }\textbf {\bibinfo {volume} {2}},\ \bibinfo {pages} {023050} (\bibinfo {year} {2020})}\BibitemShut {NoStop}%
\bibitem [{\citenamefont {Glicenstein}\ \emph {et~al.}(2021)\citenamefont {Glicenstein}, \citenamefont {Ferioli}, \citenamefont {Brossard}, \citenamefont {Sortais}, \citenamefont {Barredo}, \citenamefont {Nogrette}, \citenamefont {Ferrier-Barbut},\ and\ \citenamefont {Browaeys}}]{glicenstein2021preparation}%
  \BibitemOpen
  \bibfield  {author} {\bibinfo {author} {\bibfnamefont {A.}~\bibnamefont {Glicenstein}}, \bibinfo {author} {\bibfnamefont {G.}~\bibnamefont {Ferioli}}, \bibinfo {author} {\bibfnamefont {L.}~\bibnamefont {Brossard}}, \bibinfo {author} {\bibfnamefont {Y.~R.}\ \bibnamefont {Sortais}}, \bibinfo {author} {\bibfnamefont {D.}~\bibnamefont {Barredo}}, \bibinfo {author} {\bibfnamefont {F.}~\bibnamefont {Nogrette}}, \bibinfo {author} {\bibfnamefont {I.}~\bibnamefont {Ferrier-Barbut}},\ and\ \bibinfo {author} {\bibfnamefont {A.}~\bibnamefont {Browaeys}},\ }\bibfield  {title} {\bibinfo {title} {\href{https://journals.aps.org/pra/abstract/10.1103/PhysRevA.103.043301}{Preparation of one-dimensional chains and dense cold atomic clouds with a high numerical aperture four-lens system}},\ }\href@noop {} {\bibfield  {journal} {\bibinfo  {journal} {Physical Review A}\ }\textbf {\bibinfo {volume} {103}},\ \bibinfo {pages} {043301} (\bibinfo {year} {2021})}\BibitemShut {NoStop}%
\bibitem [{\citenamefont {Hutson}\ \emph {et~al.}(2024)\citenamefont {Hutson}, \citenamefont {Milner}, \citenamefont {Yan}, \citenamefont {Ye},\ and\ \citenamefont {Sanner}}]{hutson2024observation}%
  \BibitemOpen
  \bibfield  {author} {\bibinfo {author} {\bibfnamefont {R.~B.}\ \bibnamefont {Hutson}}, \bibinfo {author} {\bibfnamefont {W.~R.}\ \bibnamefont {Milner}}, \bibinfo {author} {\bibfnamefont {L.}~\bibnamefont {Yan}}, \bibinfo {author} {\bibfnamefont {J.}~\bibnamefont {Ye}},\ and\ \bibinfo {author} {\bibfnamefont {C.}~\bibnamefont {Sanner}},\ }\bibfield  {title} {\bibinfo {title} {\href{https://www.science.org/doi/abs/10.1126/science.adh4477}{Observation of millihertz-level cooperative Lamb shifts in an optical atomic clock}},\ }\href@noop {} {\bibfield  {journal} {\bibinfo  {journal} {Science}\ }\textbf {\bibinfo {volume} {383}},\ \bibinfo {pages} {384} (\bibinfo {year} {2024})}\BibitemShut {NoStop}%
\bibitem [{\citenamefont {Barredo}\ \emph {et~al.}(2016)\citenamefont {Barredo}, \citenamefont {De~L{\'e}s{\'e}leuc}, \citenamefont {Lienhard}, \citenamefont {Lahaye},\ and\ \citenamefont {Browaeys}}]{barredo2016atom}%
  \BibitemOpen
  \bibfield  {author} {\bibinfo {author} {\bibfnamefont {D.}~\bibnamefont {Barredo}}, \bibinfo {author} {\bibfnamefont {S.}~\bibnamefont {De~L{\'e}s{\'e}leuc}}, \bibinfo {author} {\bibfnamefont {V.}~\bibnamefont {Lienhard}}, \bibinfo {author} {\bibfnamefont {T.}~\bibnamefont {Lahaye}},\ and\ \bibinfo {author} {\bibfnamefont {A.}~\bibnamefont {Browaeys}},\ }\bibfield  {title} {\bibinfo {title} {\href{https://www.science.org/doi/abs/10.1126/science.aah3778}{An atom-by-atom assembler of defect-free arbitrary two-dimensional atomic arrays}},\ }\href@noop {} {\bibfield  {journal} {\bibinfo  {journal} {Science}\ }\textbf {\bibinfo {volume} {354}},\ \bibinfo {pages} {1021} (\bibinfo {year} {2016})}\BibitemShut {NoStop}%
\bibitem [{\citenamefont {Browaeys}\ and\ \citenamefont {Lahaye}(2020)}]{browaeys2020many}%
  \BibitemOpen
  \bibfield  {author} {\bibinfo {author} {\bibfnamefont {A.}~\bibnamefont {Browaeys}}\ and\ \bibinfo {author} {\bibfnamefont {T.}~\bibnamefont {Lahaye}},\ }\bibfield  {title} {\bibinfo {title} {\href{https://www.nature.com/articles/s41567-019-0733-z}{Many-body physics with individually controlled Rydberg atoms}},\ }\href@noop {} {\bibfield  {journal} {\bibinfo  {journal} {Nature Physics}\ }\textbf {\bibinfo {volume} {16}},\ \bibinfo {pages} {132} (\bibinfo {year} {2020})}\BibitemShut {NoStop}%
\bibitem [{\citenamefont {Labuhn}\ \emph {et~al.}(2016)\citenamefont {Labuhn}, \citenamefont {Barredo}, \citenamefont {Ravets}, \citenamefont {De~L{\'e}s{\'e}leuc}, \citenamefont {Macr{\`\i}}, \citenamefont {Lahaye},\ and\ \citenamefont {Browaeys}}]{labuhn2016tunable}%
  \BibitemOpen
  \bibfield  {author} {\bibinfo {author} {\bibfnamefont {H.}~\bibnamefont {Labuhn}}, \bibinfo {author} {\bibfnamefont {D.}~\bibnamefont {Barredo}}, \bibinfo {author} {\bibfnamefont {S.}~\bibnamefont {Ravets}}, \bibinfo {author} {\bibfnamefont {S.}~\bibnamefont {De~L{\'e}s{\'e}leuc}}, \bibinfo {author} {\bibfnamefont {T.}~\bibnamefont {Macr{\`\i}}}, \bibinfo {author} {\bibfnamefont {T.}~\bibnamefont {Lahaye}},\ and\ \bibinfo {author} {\bibfnamefont {A.}~\bibnamefont {Browaeys}},\ }\bibfield  {title} {\bibinfo {title} {\href{https://www.nature.com/articles/nature18274}{Tunable two-dimensional arrays of single Rydberg atoms for realizing quantum Ising models}},\ }\href@noop {} {\bibfield  {journal} {\bibinfo  {journal} {Nature}\ }\textbf {\bibinfo {volume} {534}},\ \bibinfo {pages} {667} (\bibinfo {year} {2016})}\BibitemShut {NoStop}%
\bibitem [{\citenamefont {Cooper}\ \emph {et~al.}(2018)\citenamefont {Cooper}, \citenamefont {Covey}, \citenamefont {Madjarov}, \citenamefont {Porsev}, \citenamefont {Safronova},\ and\ \citenamefont {Endres}}]{cooper2018alkaline}%
  \BibitemOpen
  \bibfield  {author} {\bibinfo {author} {\bibfnamefont {A.}~\bibnamefont {Cooper}}, \bibinfo {author} {\bibfnamefont {J.~P.}\ \bibnamefont {Covey}}, \bibinfo {author} {\bibfnamefont {I.~S.}\ \bibnamefont {Madjarov}}, \bibinfo {author} {\bibfnamefont {S.~G.}\ \bibnamefont {Porsev}}, \bibinfo {author} {\bibfnamefont {M.~S.}\ \bibnamefont {Safronova}},\ and\ \bibinfo {author} {\bibfnamefont {M.}~\bibnamefont {Endres}},\ }\bibfield  {title} {\bibinfo {title} {Alkaline-earth atoms in optical tweezers},\ }\href {https://doi.org/10.1103/PhysRevX.8.041055} {\bibfield  {journal} {\bibinfo  {journal} {Physical Review X}\ }\textbf {\bibinfo {volume} {8}},\ \bibinfo {pages} {041055} (\bibinfo {year} {2018})}\BibitemShut {NoStop}%
\bibitem [{\citenamefont {Norcia}\ \emph {et~al.}(2018)\citenamefont {Norcia}, \citenamefont {Young},\ and\ \citenamefont {Kaufman}}]{norcia2018microscopic}%
  \BibitemOpen
  \bibfield  {author} {\bibinfo {author} {\bibfnamefont {M.}~\bibnamefont {Norcia}}, \bibinfo {author} {\bibfnamefont {A.}~\bibnamefont {Young}},\ and\ \bibinfo {author} {\bibfnamefont {A.}~\bibnamefont {Kaufman}},\ }\bibfield  {title} {\bibinfo {title} {\href{https://journals.aps.org/prx/abstract/10.1103/PhysRevX.8.041054}{Microscopic control and detection of ultracold strontium in optical-tweezer arrays}},\ }\href@noop {} {\bibfield  {journal} {\bibinfo  {journal} {Physical Review X}\ }\textbf {\bibinfo {volume} {8}},\ \bibinfo {pages} {041054} (\bibinfo {year} {2018})}\BibitemShut {NoStop}%
\bibitem [{\citenamefont {Manetsch}\ \emph {et~al.}(2024)\citenamefont {Manetsch}, \citenamefont {Nomura}, \citenamefont {Bataille}, \citenamefont {Leung}, \citenamefont {Lv},\ and\ \citenamefont {Endres}}]{manetsch2024tweezer}%
  \BibitemOpen
  \bibfield  {author} {\bibinfo {author} {\bibfnamefont {H.~J.}\ \bibnamefont {Manetsch}}, \bibinfo {author} {\bibfnamefont {G.}~\bibnamefont {Nomura}}, \bibinfo {author} {\bibfnamefont {E.}~\bibnamefont {Bataille}}, \bibinfo {author} {\bibfnamefont {K.~H.}\ \bibnamefont {Leung}}, \bibinfo {author} {\bibfnamefont {X.}~\bibnamefont {Lv}},\ and\ \bibinfo {author} {\bibfnamefont {M.}~\bibnamefont {Endres}},\ }\bibfield  {title} {\bibinfo {title} {\href{https://arxiv.org/abs/2403.12021}{A tweezer array with 6100 highly coherent atomic qubits}},\ }\href@noop {} {\bibfield  {journal} {\bibinfo  {journal} {arXiv preprint arXiv:2403.12021}\ } (\bibinfo {year} {2024})}\BibitemShut {NoStop}%
\bibitem [{\citenamefont {Pause}\ \emph {et~al.}(2024)\citenamefont {Pause}, \citenamefont {Sturm}, \citenamefont {Mittenb{\"u}hler}, \citenamefont {Amann}, \citenamefont {Preuschoff}, \citenamefont {Sch{\"a}ffner}, \citenamefont {Schlosser},\ and\ \citenamefont {Birkl}}]{pause2024supercharged}%
  \BibitemOpen
  \bibfield  {author} {\bibinfo {author} {\bibfnamefont {L.}~\bibnamefont {Pause}}, \bibinfo {author} {\bibfnamefont {L.}~\bibnamefont {Sturm}}, \bibinfo {author} {\bibfnamefont {M.}~\bibnamefont {Mittenb{\"u}hler}}, \bibinfo {author} {\bibfnamefont {S.}~\bibnamefont {Amann}}, \bibinfo {author} {\bibfnamefont {T.}~\bibnamefont {Preuschoff}}, \bibinfo {author} {\bibfnamefont {D.}~\bibnamefont {Sch{\"a}ffner}}, \bibinfo {author} {\bibfnamefont {M.}~\bibnamefont {Schlosser}},\ and\ \bibinfo {author} {\bibfnamefont {G.}~\bibnamefont {Birkl}},\ }\bibfield  {title} {\bibinfo {title} {\href{https://arxiv.org/abs/2310.09191}{Supercharged two-dimensional tweezer array with more than 1000 atomic qubits}},\ }\href@noop {} {\bibfield  {journal} {\bibinfo  {journal} {Optica}\ }\textbf {\bibinfo {volume} {11}},\ \bibinfo {pages} {222} (\bibinfo {year} {2024})}\BibitemShut {NoStop}%
\bibitem [{\citenamefont {Verstraelen}\ \emph {et~al.}(2023)\citenamefont {Verstraelen}, \citenamefont {Huybrechts}, \citenamefont {Roscilde},\ and\ \citenamefont {Wouters}}]{verstraelen2023quantum}%
  \BibitemOpen
  \bibfield  {author} {\bibinfo {author} {\bibfnamefont {W.}~\bibnamefont {Verstraelen}}, \bibinfo {author} {\bibfnamefont {D.}~\bibnamefont {Huybrechts}}, \bibinfo {author} {\bibfnamefont {T.}~\bibnamefont {Roscilde}},\ and\ \bibinfo {author} {\bibfnamefont {M.}~\bibnamefont {Wouters}},\ }\bibfield  {title} {\bibinfo {title} {\href{https://link.aps.org/doi/10.1103/PRXQuantum.4.030304}{Quantum and classical correlations in open quantum spin lattices via truncated-cumulant trajectories}},\ }\href@noop {} {\bibfield  {journal} {\bibinfo  {journal} {PRX quantum}\ }\textbf {\bibinfo {volume} {4}},\ \bibinfo {pages} {030304} (\bibinfo {year} {2023})}\BibitemShut {NoStop}%
\bibitem [{\citenamefont {Kubo}(1962{\natexlab{a}})}]{kubo1962generalized}%
  \BibitemOpen
  \bibfield  {author} {\bibinfo {author} {\bibfnamefont {R.}~\bibnamefont {Kubo}},\ }\bibfield  {title} {\bibinfo {title} {\href{https://www.jstage.jst.go.jp/article/jpsj1946/17/7/17_7_1100/_article/-char/ja/}{Generalized cumulant expansion method}},\ }\href@noop {} {\bibfield  {journal} {\bibinfo  {journal} {Journal of the Physical Society of Japan}\ }\textbf {\bibinfo {volume} {17}},\ \bibinfo {pages} {1100} (\bibinfo {year} {1962}{\natexlab{a}})}\BibitemShut {NoStop}%
\bibitem [{\citenamefont {Rubies-Bigorda}\ \emph {et~al.}(2023{\natexlab{a}})\citenamefont {Rubies-Bigorda}, \citenamefont {Ostermann},\ and\ \citenamefont {Yelin}}]{rubies2023characterizing}%
  \BibitemOpen
  \bibfield  {author} {\bibinfo {author} {\bibfnamefont {O.}~\bibnamefont {Rubies-Bigorda}}, \bibinfo {author} {\bibfnamefont {S.}~\bibnamefont {Ostermann}},\ and\ \bibinfo {author} {\bibfnamefont {S.~F.}\ \bibnamefont {Yelin}},\ }\bibfield  {title} {\bibinfo {title} {Characterizing superradiant dynamics in atomic arrays via a cumulant expansion approach},\ }\href {https://doi.org/10.1103/PhysRevResearch.5.013091} {\bibfield  {journal} {\bibinfo  {journal} {Phys. Rev. Res.}\ }\textbf {\bibinfo {volume} {5}},\ \bibinfo {pages} {013091} (\bibinfo {year} {2023}{\natexlab{a}})}\BibitemShut {NoStop}%
\bibitem [{\citenamefont {S{\'a}nchez-Barquilla}\ \emph {et~al.}(2020)\citenamefont {S{\'a}nchez-Barquilla}, \citenamefont {Silva},\ and\ \citenamefont {Feist}}]{sanchez2020cumulant}%
  \BibitemOpen
  \bibfield  {author} {\bibinfo {author} {\bibfnamefont {M.}~\bibnamefont {S{\'a}nchez-Barquilla}}, \bibinfo {author} {\bibfnamefont {R.}~\bibnamefont {Silva}},\ and\ \bibinfo {author} {\bibfnamefont {J.}~\bibnamefont {Feist}},\ }\bibfield  {title} {\bibinfo {title} {\href{https://pubs.aip.org/aip/jcp/article/152/3/034108/198926}{Cumulant expansion for the treatment of light--matter interactions in arbitrary material structures}},\ }\href@noop {} {\bibfield  {journal} {\bibinfo  {journal} {The Journal of chemical physics}\ }\textbf {\bibinfo {volume} {152}} (\bibinfo {year} {2020})}\BibitemShut {NoStop}%
\bibitem [{\citenamefont {Masson}\ \emph {et~al.}(2024)\citenamefont {Masson}, \citenamefont {Covey}, \citenamefont {Will},\ and\ \citenamefont {Asenjo-Garcia}}]{masson2024dicke}%
  \BibitemOpen
  \bibfield  {author} {\bibinfo {author} {\bibfnamefont {S.~J.}\ \bibnamefont {Masson}}, \bibinfo {author} {\bibfnamefont {J.~P.}\ \bibnamefont {Covey}}, \bibinfo {author} {\bibfnamefont {S.}~\bibnamefont {Will}},\ and\ \bibinfo {author} {\bibfnamefont {A.}~\bibnamefont {Asenjo-Garcia}},\ }\bibfield  {title} {\bibinfo {title} {\href{https://journals.aps.org/prxquantum/abstract/10.1103/PRXQuantum.5.010344}{Dicke superradiance in ordered arrays of multilevel atoms}},\ }\href@noop {} {\bibfield  {journal} {\bibinfo  {journal} {PRX Quantum}\ }\textbf {\bibinfo {volume} {5}},\ \bibinfo {pages} {010344} (\bibinfo {year} {2024})}\BibitemShut {NoStop}%
\bibitem [{\citenamefont {Robicheaux}\ and\ \citenamefont {Suresh}(2021{\natexlab{a}})}]{robicheaux2021beyond}%
  \BibitemOpen
  \bibfield  {author} {\bibinfo {author} {\bibfnamefont {F.}~\bibnamefont {Robicheaux}}\ and\ \bibinfo {author} {\bibfnamefont {D.~A.}\ \bibnamefont {Suresh}},\ }\bibfield  {title} {\bibinfo {title} {\href{https://journals.aps.org/pra/abstract/10.1103/PhysRevA.104.023702}{Beyond lowest order mean-field theory for light interacting with atom arrays}},\ }\href@noop {} {\bibfield  {journal} {\bibinfo  {journal} {Physical Review A}\ }\textbf {\bibinfo {volume} {104}},\ \bibinfo {pages} {023702} (\bibinfo {year} {2021}{\natexlab{a}})}\BibitemShut {NoStop}%
\bibitem [{\citenamefont {Kr{\"a}mer}\ and\ \citenamefont {Ritsch}(2015)}]{kramer2015generalized}%
  \BibitemOpen
  \bibfield  {author} {\bibinfo {author} {\bibfnamefont {S.}~\bibnamefont {Kr{\"a}mer}}\ and\ \bibinfo {author} {\bibfnamefont {H.}~\bibnamefont {Ritsch}},\ }\bibfield  {title} {\bibinfo {title} {\href{https://link.springer.com/article/10.1140/epjd/e2015-60266-5}{Generalized mean-field approach to simulate the dynamics of large open spin ensembles with long range interactions}},\ }\href {https://doi.org/10.1140/epjd/e2015-60266-5} {\bibfield  {journal} {\bibinfo  {journal} {The European Physical Journal D}\ }\textbf {\bibinfo {volume} {69}},\ \bibinfo {pages} {282} (\bibinfo {year} {2015})}\BibitemShut {NoStop}%
\bibitem [{\citenamefont {Rubies-Bigorda}\ \emph {et~al.}(2023{\natexlab{b}})\citenamefont {Rubies-Bigorda}, \citenamefont {Ostermann},\ and\ \citenamefont {Yelin}}]{Rubies_subradiance_cumulant}%
  \BibitemOpen
  \bibfield  {author} {\bibinfo {author} {\bibfnamefont {O.}~\bibnamefont {Rubies-Bigorda}}, \bibinfo {author} {\bibfnamefont {S.}~\bibnamefont {Ostermann}},\ and\ \bibinfo {author} {\bibfnamefont {S.~F.}\ \bibnamefont {Yelin}},\ }\bibfield  {title} {\bibinfo {title} {Dynamic population of multiexcitation subradiant states in incoherently excited atomic arrays},\ }\href {https://doi.org/10.1103/PhysRevA.107.L051701} {\bibfield  {journal} {\bibinfo  {journal} {Phys. Rev. A}\ }\textbf {\bibinfo {volume} {107}},\ \bibinfo {pages} {L051701} (\bibinfo {year} {2023}{\natexlab{b}})}\BibitemShut {NoStop}%
\bibitem [{\citenamefont {Dicke}(1954)}]{Dicke_originalpaper}%
  \BibitemOpen
  \bibfield  {author} {\bibinfo {author} {\bibfnamefont {R.~H.}\ \bibnamefont {Dicke}},\ }\bibfield  {title} {\bibinfo {title} {Coherence in spontaneous radiation processes},\ }\href {https://doi.org/10.1103/PhysRev.93.99} {\bibfield  {journal} {\bibinfo  {journal} {Phys. Rev.}\ }\textbf {\bibinfo {volume} {93}},\ \bibinfo {pages} {99} (\bibinfo {year} {1954})}\BibitemShut {NoStop}%
\bibitem [{\citenamefont {Asenjo-Garcia~et al.}(2017)}]{asenjo2017exponential}%
  \BibitemOpen
  \bibfield  {author} {\bibinfo {author} {\bibfnamefont {A.}~\bibnamefont {Asenjo-Garcia~et al.}},\ }\bibfield  {title} {\bibinfo {title} {\href{https://link.aps.org/doi/10.1103/PhysRevX.7.031024}{Exponential Improvement in Photon Storage Fidelities Using Subradiance and ``Selective Radiance'' in Atomic Arrays}},\ }\href {https://doi.org/10.1103/PhysRevX.7.031024} {\bibfield  {journal} {\bibinfo  {journal} {Phys. Rev. X}\ }\textbf {\bibinfo {volume} {7}},\ \bibinfo {pages} {031024} (\bibinfo {year} {2017})}\BibitemShut {NoStop}%
\bibitem [{\citenamefont {W{\"u}rthner}\ \emph {et~al.}(2011)\citenamefont {W{\"u}rthner}, \citenamefont {Kaiser},\ and\ \citenamefont {Saha-M{\"o}ller}}]{wurthner2011j}%
  \BibitemOpen
  \bibfield  {author} {\bibinfo {author} {\bibfnamefont {F.}~\bibnamefont {W{\"u}rthner}}, \bibinfo {author} {\bibfnamefont {T.~E.}\ \bibnamefont {Kaiser}},\ and\ \bibinfo {author} {\bibfnamefont {C.~R.}\ \bibnamefont {Saha-M{\"o}ller}},\ }\bibfield  {title} {\bibinfo {title} {\href{https://onlinelibrary.wiley.com/doi/abs/10.1002/anie.201002307}{J-aggregates: from serendipitous discovery to supramolecular engineering of functional dye materials}},\ }\href@noop {} {\bibfield  {journal} {\bibinfo  {journal} {Angewandte Chemie International Edition}\ }\textbf {\bibinfo {volume} {50}},\ \bibinfo {pages} {3376} (\bibinfo {year} {2011})}\BibitemShut {NoStop}%
\bibitem [{\citenamefont {Pennetta}\ \emph {et~al.}(2022)\citenamefont {Pennetta}, \citenamefont {Blaha}, \citenamefont {Johnson}, \citenamefont {Lechner}, \citenamefont {Schneeweiss}, \citenamefont {Volz},\ and\ \citenamefont {Rauschenbeutel}}]{pennetta2022collective}%
  \BibitemOpen
  \bibfield  {author} {\bibinfo {author} {\bibfnamefont {R.}~\bibnamefont {Pennetta}}, \bibinfo {author} {\bibfnamefont {M.}~\bibnamefont {Blaha}}, \bibinfo {author} {\bibfnamefont {A.}~\bibnamefont {Johnson}}, \bibinfo {author} {\bibfnamefont {D.}~\bibnamefont {Lechner}}, \bibinfo {author} {\bibfnamefont {P.}~\bibnamefont {Schneeweiss}}, \bibinfo {author} {\bibfnamefont {J.}~\bibnamefont {Volz}},\ and\ \bibinfo {author} {\bibfnamefont {A.}~\bibnamefont {Rauschenbeutel}},\ }\bibfield  {title} {\bibinfo {title} {\href{https://journals.aps.org/prl/abstract/10.1103/PhysRevLett.128.073601}{Collective radiative dynamics of an ensemble of cold atoms coupled to an optical waveguide}},\ }\href@noop {} {\bibfield  {journal} {\bibinfo  {journal} {Physical Review Letters}\ }\textbf {\bibinfo {volume} {128}},\ \bibinfo {pages} {073601} (\bibinfo {year} {2022})}\BibitemShut {NoStop}%
\bibitem [{\citenamefont {Tebbenjohanns}\ \emph {et~al.}(2024)\citenamefont {Tebbenjohanns}, \citenamefont {Mink}, \citenamefont {Bach}, \citenamefont {Rauschenbeutel},\ and\ \citenamefont {Fleischhauer}}]{tebbenjohanns2024predicting}%
  \BibitemOpen
  \bibfield  {author} {\bibinfo {author} {\bibfnamefont {F.}~\bibnamefont {Tebbenjohanns}}, \bibinfo {author} {\bibfnamefont {C.~D.}\ \bibnamefont {Mink}}, \bibinfo {author} {\bibfnamefont {C.}~\bibnamefont {Bach}}, \bibinfo {author} {\bibfnamefont {A.}~\bibnamefont {Rauschenbeutel}},\ and\ \bibinfo {author} {\bibfnamefont {M.}~\bibnamefont {Fleischhauer}},\ }\bibfield  {title} {\bibinfo {title} {\href{https://arxiv.org/abs/2407.02154}{Predicting correlations in superradiant emission from a cascaded quantum system}},\ }\href@noop {} {\bibfield  {journal} {\bibinfo  {journal} {arXiv preprint arXiv:2407.02154}\ } (\bibinfo {year} {2024})}\BibitemShut {NoStop}%
\bibitem [{\citenamefont {Liedl}\ \emph {et~al.}(2024)\citenamefont {Liedl}, \citenamefont {Tebbenjohanns}, \citenamefont {Bach}, \citenamefont {Pucher}, \citenamefont {Rauschenbeutel},\ and\ \citenamefont {Schneeweiss}}]{liedl2024observation}%
  \BibitemOpen
  \bibfield  {author} {\bibinfo {author} {\bibfnamefont {C.}~\bibnamefont {Liedl}}, \bibinfo {author} {\bibfnamefont {F.}~\bibnamefont {Tebbenjohanns}}, \bibinfo {author} {\bibfnamefont {C.}~\bibnamefont {Bach}}, \bibinfo {author} {\bibfnamefont {S.}~\bibnamefont {Pucher}}, \bibinfo {author} {\bibfnamefont {A.}~\bibnamefont {Rauschenbeutel}},\ and\ \bibinfo {author} {\bibfnamefont {P.}~\bibnamefont {Schneeweiss}},\ }\bibfield  {title} {\bibinfo {title} {\href{https://journals.aps.org/prx/abstract/10.1103/PhysRevX.14.011020}{Observation of superradiant bursts in a cascaded quantum system}},\ }\href@noop {} {\bibfield  {journal} {\bibinfo  {journal} {Physical Review X}\ }\textbf {\bibinfo {volume} {14}},\ \bibinfo {pages} {011020} (\bibinfo {year} {2024})}\BibitemShut {NoStop}%
\bibitem [{\citenamefont {Goban}\ \emph {et~al.}(2015)\citenamefont {Goban}, \citenamefont {Hung}, \citenamefont {Hood}, \citenamefont {Yu}, \citenamefont {Muniz}, \citenamefont {Painter},\ and\ \citenamefont {Kimble}}]{goban2015superradiance}%
  \BibitemOpen
  \bibfield  {author} {\bibinfo {author} {\bibfnamefont {A.}~\bibnamefont {Goban}}, \bibinfo {author} {\bibfnamefont {C.-L.}\ \bibnamefont {Hung}}, \bibinfo {author} {\bibfnamefont {J.}~\bibnamefont {Hood}}, \bibinfo {author} {\bibfnamefont {S.-P.}\ \bibnamefont {Yu}}, \bibinfo {author} {\bibfnamefont {J.}~\bibnamefont {Muniz}}, \bibinfo {author} {\bibfnamefont {O.}~\bibnamefont {Painter}},\ and\ \bibinfo {author} {\bibfnamefont {H.}~\bibnamefont {Kimble}},\ }\bibfield  {title} {\bibinfo {title} {\href{https://journals.aps.org/prl/abstract/10.1103/PhysRevLett.115.063601}{Superradiance for atoms trapped along a photonic crystal waveguide}},\ }\href@noop {} {\bibfield  {journal} {\bibinfo  {journal} {Physical review letters}\ }\textbf {\bibinfo {volume} {115}},\ \bibinfo {pages} {063601} (\bibinfo {year} {2015})}\BibitemShut {NoStop}%
\bibitem [{\citenamefont {Li}\ \emph {et~al.}(2023)\citenamefont {Li}, \citenamefont {Marino}, \citenamefont {Chang},\ and\ \citenamefont {Flebus}}]{li2023solid}%
  \BibitemOpen
  \bibfield  {author} {\bibinfo {author} {\bibfnamefont {X.}~\bibnamefont {Li}}, \bibinfo {author} {\bibfnamefont {J.}~\bibnamefont {Marino}}, \bibinfo {author} {\bibfnamefont {D.~E.}\ \bibnamefont {Chang}},\ and\ \bibinfo {author} {\bibfnamefont {B.}~\bibnamefont {Flebus}},\ }\bibfield  {title} {\bibinfo {title} {\href{https://arxiv.org/abs/2309.08991}{A solid-state platform for cooperative quantum phenomena}},\ }\href@noop {} {\bibfield  {journal} {\bibinfo  {journal} {arXiv preprint arXiv:2309.08991}\ } (\bibinfo {year} {2023})}\BibitemShut {NoStop}%
\bibitem [{\citenamefont {Rain{\`o}}\ \emph {et~al.}(2018)\citenamefont {Rain{\`o}}, \citenamefont {Becker}, \citenamefont {Bodnarchuk}, \citenamefont {Mahrt}, \citenamefont {Kovalenko},\ and\ \citenamefont {St{\"o}ferle}}]{raino2018superfluorescence}%
  \BibitemOpen
  \bibfield  {author} {\bibinfo {author} {\bibfnamefont {G.}~\bibnamefont {Rain{\`o}}}, \bibinfo {author} {\bibfnamefont {M.~A.}\ \bibnamefont {Becker}}, \bibinfo {author} {\bibfnamefont {M.~I.}\ \bibnamefont {Bodnarchuk}}, \bibinfo {author} {\bibfnamefont {R.~F.}\ \bibnamefont {Mahrt}}, \bibinfo {author} {\bibfnamefont {M.~V.}\ \bibnamefont {Kovalenko}},\ and\ \bibinfo {author} {\bibfnamefont {T.}~\bibnamefont {St{\"o}ferle}},\ }\bibfield  {title} {\bibinfo {title} {\href{https://www.nature.com/articles/s41586-018-0683-0}{Superfluorescence from lead halide perovskite quantum dot superlattices}},\ }\href@noop {} {\bibfield  {journal} {\bibinfo  {journal} {Nature}\ }\textbf {\bibinfo {volume} {563}},\ \bibinfo {pages} {671} (\bibinfo {year} {2018})}\BibitemShut {NoStop}%
\bibitem [{\citenamefont {Lehmberg}(1970{\natexlab{a}})}]{lehmberg1970radiation}%
  \BibitemOpen
  \bibfield  {author} {\bibinfo {author} {\bibfnamefont {R.}~\bibnamefont {Lehmberg}},\ }\bibfield  {title} {\bibinfo {title} {\href{https://journals.aps.org/pra/abstract/10.1103/PhysRevA.2.883}{Radiation from an N-atom system. I. General formalism}},\ }\href@noop {} {\bibfield  {journal} {\bibinfo  {journal} {Physical Review A}\ }\textbf {\bibinfo {volume} {2}},\ \bibinfo {pages} {883} (\bibinfo {year} {1970}{\natexlab{a}})}\BibitemShut {NoStop}%
\bibitem [{\citenamefont {Novotny}\ and\ \citenamefont {Hecht}(2012)}]{novotny2012principles}%
  \BibitemOpen
  \bibfield  {author} {\bibinfo {author} {\bibfnamefont {L.}~\bibnamefont {Novotny}}\ and\ \bibinfo {author} {\bibfnamefont {B.}~\bibnamefont {Hecht}},\ }\href@noop {} {\emph {\bibinfo {title} {\href{https://www.cambridge.org/core/books/principles-of-nanooptics/E884E5F4AA76DF179A1ECFDF77436452}{Principles of nano-optics}}}}\ (\bibinfo  {publisher} {Cambridge university press},\ \bibinfo {year} {2012})\BibitemShut {NoStop}%
\bibitem [{\citenamefont {Moreno-Cardoner}\ \emph {et~al.}(2019)\citenamefont {Moreno-Cardoner}, \citenamefont {Plankensteiner}, \citenamefont {Ostermann}, \citenamefont {Chang},\ and\ \citenamefont {Ritsch}}]{moreno2019subradiance}%
  \BibitemOpen
  \bibfield  {author} {\bibinfo {author} {\bibfnamefont {M.}~\bibnamefont {Moreno-Cardoner}}, \bibinfo {author} {\bibfnamefont {D.}~\bibnamefont {Plankensteiner}}, \bibinfo {author} {\bibfnamefont {L.}~\bibnamefont {Ostermann}}, \bibinfo {author} {\bibfnamefont {D.~E.}\ \bibnamefont {Chang}},\ and\ \bibinfo {author} {\bibfnamefont {H.}~\bibnamefont {Ritsch}},\ }\bibfield  {title} {\bibinfo {title} {\href{https://journals.aps.org/pra/abstract/10.1103/PhysRevA.100.023806}{Subradiance-enhanced excitation transfer between dipole-coupled nanorings of quantum emitters}},\ }\href@noop {} {\bibfield  {journal} {\bibinfo  {journal} {Physical Review A}\ }\textbf {\bibinfo {volume} {100}},\ \bibinfo {pages} {023806} (\bibinfo {year} {2019})}\BibitemShut {NoStop}%
\bibitem [{\citenamefont {Lehmberg}(1970{\natexlab{b}})}]{Lehmberg1970_1}%
  \BibitemOpen
  \bibfield  {author} {\bibinfo {author} {\bibfnamefont {R.~H.}\ \bibnamefont {Lehmberg}},\ }\bibfield  {title} {\bibinfo {title} {Radiation from an $n$-atom system. i. general formalism},\ }\href {https://doi.org/10.1103/PhysRevA.2.883} {\bibfield  {journal} {\bibinfo  {journal} {Phys. Rev. A}\ }\textbf {\bibinfo {volume} {2}},\ \bibinfo {pages} {883} (\bibinfo {year} {1970}{\natexlab{b}})}\BibitemShut {NoStop}%
\bibitem [{\citenamefont {Kubo}(1962{\natexlab{b}})}]{Cumulant_Kubo}%
  \BibitemOpen
  \bibfield  {author} {\bibinfo {author} {\bibfnamefont {R.}~\bibnamefont {Kubo}},\ }\bibfield  {title} {\bibinfo {title} {\href{https://www.jstage.jst.go.jp/article/jpsj1946/17/7/17_7_1100/_article/-char/ja/}{Generalized Cumulant Expansion Method}},\ }\href {https://doi.org/10.1143/JPSJ.17.1100} {\bibfield  {journal} {\bibinfo  {journal} {Journal of the Physical Society of Japan}\ }\textbf {\bibinfo {volume} {17}},\ \bibinfo {pages} {1100} (\bibinfo {year} {1962}{\natexlab{b}})}\BibitemShut {NoStop}%
\bibitem [{\citenamefont {Robicheaux}\ and\ \citenamefont {Suresh}(2021{\natexlab{b}})}]{Robicheaux_cumulants}%
  \BibitemOpen
  \bibfield  {author} {\bibinfo {author} {\bibfnamefont {F.}~\bibnamefont {Robicheaux}}\ and\ \bibinfo {author} {\bibfnamefont {D.~A.}\ \bibnamefont {Suresh}},\ }\bibfield  {title} {\bibinfo {title} {Beyond lowest order mean-field theory for light interacting with atom arrays},\ }\href {https://doi.org/10.1103/PhysRevA.104.023702} {\bibfield  {journal} {\bibinfo  {journal} {Phys. Rev. A}\ }\textbf {\bibinfo {volume} {104}},\ \bibinfo {pages} {023702} (\bibinfo {year} {2021}{\natexlab{b}})}\BibitemShut {NoStop}%
\bibitem [{\citenamefont {Dum}\ \emph {et~al.}(1992)\citenamefont {Dum}, \citenamefont {Zoller},\ and\ \citenamefont {Ritsch}}]{dum1992monte}%
  \BibitemOpen
  \bibfield  {author} {\bibinfo {author} {\bibfnamefont {R.}~\bibnamefont {Dum}}, \bibinfo {author} {\bibfnamefont {P.}~\bibnamefont {Zoller}},\ and\ \bibinfo {author} {\bibfnamefont {H.}~\bibnamefont {Ritsch}},\ }\bibfield  {title} {\bibinfo {title} {\href{https://journals.aps.org/pra/abstract/10.1103/PhysRevA.45.4879}{Monte Carlo simulation of the atomic master equation for spontaneous emission}},\ }\href@noop {} {\bibfield  {journal} {\bibinfo  {journal} {Physical Review A}\ }\textbf {\bibinfo {volume} {45}},\ \bibinfo {pages} {4879} (\bibinfo {year} {1992})}\BibitemShut {NoStop}%
\bibitem [{\citenamefont {Mølmer}\ \emph {et~al.}(1993)\citenamefont {Mølmer}, \citenamefont {Castin},\ and\ \citenamefont {Dalibard}}]{molmer_monte_1993}%
  \BibitemOpen
  \bibfield  {author} {\bibinfo {author} {\bibfnamefont {K.}~\bibnamefont {Mølmer}}, \bibinfo {author} {\bibfnamefont {Y.}~\bibnamefont {Castin}},\ and\ \bibinfo {author} {\bibfnamefont {J.}~\bibnamefont {Dalibard}},\ }\bibfield  {title} {\bibinfo {title} {Monte {Carlo} wave-function method in quantum optics},\ }\href {https://doi.org/10.1364/JOSAB.10.000524} {\bibfield  {journal} {\bibinfo  {journal} {Journal of the Optical Society of America B}\ }\textbf {\bibinfo {volume} {10}},\ \bibinfo {pages} {524} (\bibinfo {year} {1993})}\BibitemShut {NoStop}%
\bibitem [{\citenamefont {Bezanson}\ \emph {et~al.}(2017)\citenamefont {Bezanson}, \citenamefont {Edelman}, \citenamefont {Karpinski},\ and\ \citenamefont {Shah}}]{bezanson2017julia}%
  \BibitemOpen
  \bibfield  {author} {\bibinfo {author} {\bibfnamefont {J.}~\bibnamefont {Bezanson}}, \bibinfo {author} {\bibfnamefont {A.}~\bibnamefont {Edelman}}, \bibinfo {author} {\bibfnamefont {S.}~\bibnamefont {Karpinski}},\ and\ \bibinfo {author} {\bibfnamefont {V.~B.}\ \bibnamefont {Shah}},\ }\bibfield  {title} {\bibinfo {title} {Julia: A fresh approach to numerical computing},\ }\href@noop {} {\bibfield  {journal} {\bibinfo  {journal} {SIAM review}\ }\textbf {\bibinfo {volume} {59}},\ \bibinfo {pages} {65} (\bibinfo {year} {2017})}\BibitemShut {NoStop}%
\bibitem [{\citenamefont {Rackauckas}\ and\ \citenamefont {Nie}(2017)}]{rackauckas2017differentialequations}%
  \BibitemOpen
  \bibfield  {author} {\bibinfo {author} {\bibfnamefont {C.}~\bibnamefont {Rackauckas}}\ and\ \bibinfo {author} {\bibfnamefont {Q.}~\bibnamefont {Nie}},\ }\bibfield  {title} {\bibinfo {title} {Differentialequations. jl--a performant and feature-rich ecosystem for solving differential equations in julia},\ }\href@noop {} {\bibfield  {journal} {\bibinfo  {journal} {Journal of open research software}\ }\textbf {\bibinfo {volume} {5}},\ \bibinfo {pages} {15} (\bibinfo {year} {2017})}\BibitemShut {NoStop}%
\bibitem [{\citenamefont {Robicheaux}(2021)}]{Robicheaux_superradiance}%
  \BibitemOpen
  \bibfield  {author} {\bibinfo {author} {\bibfnamefont {F.}~\bibnamefont {Robicheaux}},\ }\bibfield  {title} {\bibinfo {title} {Theoretical study of early-time superradiance for atom clouds and arrays},\ }\href {https://doi.org/10.1103/PhysRevA.104.063706} {\bibfield  {journal} {\bibinfo  {journal} {Phys. Rev. A}\ }\textbf {\bibinfo {volume} {104}},\ \bibinfo {pages} {063706} (\bibinfo {year} {2021})}\BibitemShut {NoStop}%
\bibitem [{\citenamefont {Baier}\ \emph {et~al.}(2016)\citenamefont {Baier}, \citenamefont {Mark}, \citenamefont {Petter}, \citenamefont {Aikawa}, \citenamefont {Chomaz}, \citenamefont {Cai}, \citenamefont {Baranov}, \citenamefont {Zoller},\ and\ \citenamefont {Ferlaino}}]{baier2016extended}%
  \BibitemOpen
  \bibfield  {author} {\bibinfo {author} {\bibfnamefont {S.}~\bibnamefont {Baier}}, \bibinfo {author} {\bibfnamefont {M.~J.}\ \bibnamefont {Mark}}, \bibinfo {author} {\bibfnamefont {D.}~\bibnamefont {Petter}}, \bibinfo {author} {\bibfnamefont {K.}~\bibnamefont {Aikawa}}, \bibinfo {author} {\bibfnamefont {L.}~\bibnamefont {Chomaz}}, \bibinfo {author} {\bibfnamefont {Z.}~\bibnamefont {Cai}}, \bibinfo {author} {\bibfnamefont {M.}~\bibnamefont {Baranov}}, \bibinfo {author} {\bibfnamefont {P.}~\bibnamefont {Zoller}},\ and\ \bibinfo {author} {\bibfnamefont {F.}~\bibnamefont {Ferlaino}},\ }\bibfield  {title} {\bibinfo {title} {\href{https://www.science.org/doi/abs/10.1126/science.aac9812}{Extended Bose-Hubbard models with ultracold magnetic atoms}},\ }\href@noop {} {\bibfield  {journal} {\bibinfo  {journal} {Science}\ }\textbf {\bibinfo {volume} {352}},\ \bibinfo {pages} {201} (\bibinfo {year} {2016})}\BibitemShut {NoStop}%
\bibitem [{\citenamefont {Ostermann}\ \emph {et~al.}(2024)\citenamefont {Ostermann}, \citenamefont {Rubies-Bigorda}, \citenamefont {Zhang},\ and\ \citenamefont {Yelin}}]{Ostermann_DDM}%
  \BibitemOpen
  \bibfield  {author} {\bibinfo {author} {\bibfnamefont {S.}~\bibnamefont {Ostermann}}, \bibinfo {author} {\bibfnamefont {O.}~\bibnamefont {Rubies-Bigorda}}, \bibinfo {author} {\bibfnamefont {V.}~\bibnamefont {Zhang}},\ and\ \bibinfo {author} {\bibfnamefont {S.~F.}\ \bibnamefont {Yelin}},\ }\bibfield  {title} {\bibinfo {title} {Breakdown of steady-state superradiance in extended driven atomic arrays},\ }\href {https://doi.org/10.1103/PhysRevResearch.6.023206} {\bibfield  {journal} {\bibinfo  {journal} {Phys. Rev. Res.}\ }\textbf {\bibinfo {volume} {6}},\ \bibinfo {pages} {023206} (\bibinfo {year} {2024})}\BibitemShut {NoStop}%
\bibitem [{\citenamefont {Sheremet}\ \emph {et~al.}(2023)\citenamefont {Sheremet}, \citenamefont {Petrov}, \citenamefont {Iorsh}, \citenamefont {Poshakinskiy},\ and\ \citenamefont {Poddubny}}]{sheremet2023waveguide}%
  \BibitemOpen
  \bibfield  {author} {\bibinfo {author} {\bibfnamefont {A.~S.}\ \bibnamefont {Sheremet}}, \bibinfo {author} {\bibfnamefont {M.~I.}\ \bibnamefont {Petrov}}, \bibinfo {author} {\bibfnamefont {I.~V.}\ \bibnamefont {Iorsh}}, \bibinfo {author} {\bibfnamefont {A.~V.}\ \bibnamefont {Poshakinskiy}},\ and\ \bibinfo {author} {\bibfnamefont {A.~N.}\ \bibnamefont {Poddubny}},\ }\bibfield  {title} {\bibinfo {title} {\href{https://journals.aps.org/rmp/abstract/10.1103/RevModPhys.95.015002}{Waveguide quantum electrodynamics: Collective radiance and photon-photon correlations}},\ }\href@noop {} {\bibfield  {journal} {\bibinfo  {journal} {Reviews of Modern Physics}\ }\textbf {\bibinfo {volume} {95}},\ \bibinfo {pages} {015002} (\bibinfo {year} {2023})}\BibitemShut {NoStop}%
\bibitem [{\citenamefont {Kira}\ and\ \citenamefont {Koch}(2008)}]{kira2008cluster}%
  \BibitemOpen
  \bibfield  {author} {\bibinfo {author} {\bibfnamefont {M.}~\bibnamefont {Kira}}\ and\ \bibinfo {author} {\bibfnamefont {S.}~\bibnamefont {Koch}},\ }\bibfield  {title} {\bibinfo {title} {\href{https://journals.aps.org/pra/abstract/10.1103/PhysRevA.78.022102}{Cluster-expansion representation in quantum optics}},\ }\href@noop {} {\bibfield  {journal} {\bibinfo  {journal} {Physical Review A—Atomic, Molecular, and Optical Physics}\ }\textbf {\bibinfo {volume} {78}},\ \bibinfo {pages} {022102} (\bibinfo {year} {2008})}\BibitemShut {NoStop}%
\bibitem [{\citenamefont {Moreno-Cardoner}\ \emph {et~al.}(2022)\citenamefont {Moreno-Cardoner}, \citenamefont {Holzinger},\ and\ \citenamefont {Ritsch}}]{moreno2022efficient}%
  \BibitemOpen
  \bibfield  {author} {\bibinfo {author} {\bibfnamefont {M.}~\bibnamefont {Moreno-Cardoner}}, \bibinfo {author} {\bibfnamefont {R.}~\bibnamefont {Holzinger}},\ and\ \bibinfo {author} {\bibfnamefont {H.}~\bibnamefont {Ritsch}},\ }\bibfield  {title} {\bibinfo {title} {\href{https://opg.optica.org/oe/fulltext.cfm?uri=oe-30-7-10779&id=470528}{Efficient nano-photonic antennas based on dark states in quantum emitter rings}},\ }\href@noop {} {\bibfield  {journal} {\bibinfo  {journal} {Optics Express}\ }\textbf {\bibinfo {volume} {30}},\ \bibinfo {pages} {10779} (\bibinfo {year} {2022})}\BibitemShut {NoStop}%
\end{thebibliography}%

\pagebreak

\clearpage
\onecolumngrid
\renewcommand{\thesection}{\Alph{section}}
\setcounter{section}{0}

\begin{center}

\newcommand{\beginsupplement}{%
        \setcounter{table}{0}
        \renewcommand{\thetable}{S\arabic{table}}%
        \setcounter{figure}{0}
        \renewcommand{\thefigure}{S\arabic{figure}}%
     }
\textbf{\large Supplemental Material}
\end{center}
\newcommand{\beginsupplement}{%
        \setcounter{table}{0}
        \renewcommand{\thetable}{S\arabic{table}}%
        \setcounter{figure}{0}
        \renewcommand{\thefigure}{S\arabic{figure}}%
     }
\setcounter{equation}{0}
\setcounter{figure}{0}
\setcounter{table}{0}   
\setcounter{page}{1}
\makeatletter
\renewcommand{\theequation}{S\arabic{equation}}
\renewcommand{\thefigure}{S\arabic{figure}}
\renewcommand{\bibnumfmt}[1]{[S#1]}
\renewcommand{\citenumfont}[1]{S#1}
\vspace{0.8 in}
\newcommand{\D}{\Delta}
\newcommand{\tD}{\tilde{\Delta}}
\newcommand{\K}{K_{PP}}
\newcommand{\bn}{\bar{n}_P}
\newcommand{\G}{\Gamma}
\newcommand{\LH}{\underset{L}{H}}
\newcommand{\HL}{\underset{H}{L}}
\vspace{-1in}

\section{Green's function}
\label{SI: Green}

The coherent and dissipative interaction rates in the main text (Eq.~(\ref{dipole-interaction}) between emitters $n$ and $m$ are given by
\begin{align}
   J_{nm} - \frac{i\Gamma_{nm}}{2} = -\frac{3 \pi \gamma_0}{\omega_0} \mathbf{d}^\dagger \cdot \mathbf{G}(\mathbf{r}_{nm},\omega_0) \cdot \mathbf{d},
\end{align}
where $\mathbf{d}$ is the transition dipole moment matrix element and $\mathbf{r}_{nm} = \mathbf{r}_n-\mathbf{r}_m$ is the connecting vector between emitters $n$ and $m$.
The Green's tensor $\mathbf{G}(\mathbf{r}_{nm},\omega_0)$ is the propagator of the electromagnetic
field between emitter positions $\boldsymbol{r}_n$ and $\boldsymbol{r}_m$, and reads
\begin{align}
\mathbf{G}(\mathbf{r}_{nm},\omega_0) = \frac{e^{i k_0 r_{nm}}}{4\pi k_0^2 r_{nm}^3} \bigg[ \left( k_0^2 r_{nm}^2 + ik_0 r_{nm} -1 \right) \mathbb{1}  +  \left(-k_0^2 r_{nm}^2 - 3i k_0 r_{nm} + 3 \right) \frac{\mathbf{r}_{nm} \otimes \mathbf{r}_{nm}}{r_{nm}^2} \bigg],
\end{align}
with $r_{nm} = |\mathbf{r}_{nm}|$. The spontaneos decay rate is equal for all atoms, $\Gamma_{nn} = \gamma_0$, while the self-energy term $J_{nn}$ is set to zero, as it simply leads to renormalization of the transition energies $\omega_0$. 

\section{Collective spin operators for finite arrays and open boundaries} \label{appendix:finite}
\begin{figure}[ht!]
    \centering
    \includegraphics[width = 0.85\columnwidth]{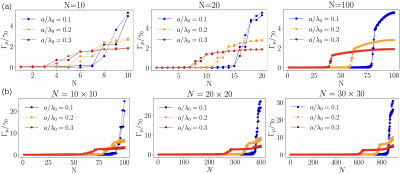}
    \caption{Comparison of the collective decay rates between numerical diagonalization (dark colors) and the expressions in Eq.~\ref{supp:rates}(a) (bright colors and dashed-lines) for various array spacings $a$. The decay rates are sorted from most subradiant (left) to most (super)radiant (right). (a) For a linear chain of emitters. (b) For a square array. In all plots: $\boldsymbol{d} = (1,i,0)^T$.}
    \label{fig:supp1}
\end{figure}

For ordered arrays with finite emitter numbers and open boundaries, the expressions for the collective eigenvalues
\begin{subequations} \label{supp:rates}
\begin{align} 
    \Gamma_{\boldsymbol{\mu}} &=  \frac{1}{N} \sum_{{n},{m}}^N \mathrm{exp}\Big( i\frac{2\pi}{N_\mathrm{1D} a}  \boldsymbol{\mu} \cdot (\boldsymbol{r}_n-\boldsymbol{r}_m)  \Big) \ \Gamma_{{n} {m}} \\
    J_{\boldsymbol{\mu}} &= \frac{1}{N} \sum_{{n}\neq{m}}^N \mathrm{exp}\Big( i\frac{2\pi}{N_\mathrm{1D} a}  \boldsymbol{\mu} \cdot (\boldsymbol{r}_n-\boldsymbol{r}_m)  \Big) \ J_{{n} {m}},
    \end{align}
\end{subequations}
are only approximations and are exact only for infinite arrays or blosed boundaries such as ring geometries~\cite{asenjo2017exponential}. To quantify the approximation, we compare the collective decay rates with the exact numerical diagonalization of the decay matrix $\boldsymbol{\Gamma}$ in Fig.~\ref{fig:supp1} where the rates are ordered from most subradiant to superradiant. Fast convergence to the numerically exact eigenvalues is observed for the one-dimensional chain geometry, while for the two-dimensional square array, slower convergence is observed, but with overall qualitative agreement. We note, that eventhough there is noticable difference in the most superradiant decay rate for the square array in Fig.~\ref{fig:supp1}(b), the resulting photon emission rate observed in the time dynamics exhibits excellent agreement.
\subsection*{Arbitrary geometries of emitter ensembles}
The discussion has been focused on ordered emitter arrays exploiting the translation symmetries along all directions, but the equations of motion derived in Section~\ref{supp:cumulant-collective}, can generally be derived for arbitrary emitter ensembles, in the absence of any symmetries.

Starting from the master equation in Eq.\ref{master}, we numerically diagonalize the decay matrix $\boldsymbol{\Gamma}$ to obtain the collective decay rates $\Gamma_{{\mu}}$ and collective spin operators $\hat{O}_{{\mu}}$ as the associated eigenvectors, with $\mu = 1,\cdots,N$. The collective spin operators are now generally superpositions of the individual spins, $\hat{O}_{{\mu}} = \sum_n \alpha_{\mu,n} \hat{\sigma}_n $ and the complex coefficients $\alpha_{\mu,n}$ depend on the specific geometry, with the property $N^{-1}\sum_{n} \alpha^*_{\mu,n} \alpha_{\nu,n} = \delta_{\mu \nu}$~\cite{masson2024dicke}. Now, instead of the commutation relations in Eq.~\ref{commutation1}, we obtain
\begin{subequations} \label{supp:comm}
\begin{align} 
    \Big[\hat{O}_{{\mu}}^\dagger,\hat{O}_{{\nu}} \Big] &= 2 \hat{O}_{{\mu}-{\nu}}^{ee} - \hat{\mathds{1}} \delta_{_{{\mu \nu}}} \\
\Big[\hat{O}^\dagger_{{\mu}},\hat{O}_{{\nu}}^{ee} \Big]  &= - \frac{1}{N} \hat{O}^\dagger_{{\mu+\nu}} \\
\Big[\hat{O}_{{\mu}},\hat{O}_{{\nu}}^{ee} \Big]  &= \frac{1}{N} \hat{O}_{{\mu-\nu}}.
\end{align}
\end{subequations}
For instance, Eq.~\ref{supp:comm}(a) is evaluated as
\begin{equation}
    [\hat{O}_\mu^\dagger,\hat{O}_\nu] = \frac{1}{N} \sum_{nm} \alpha_{\mu,n}^* \alpha_{\nu,m} [\hat{\sigma}_n^\dagger,\hat{\sigma}_m] = \frac{1}{N} \sum_{i} \alpha_{\mu,n}^* \alpha_{\nu,n}(2\hat{\sigma}_n^{ee}-1) = 2\hat{O}^{ee}_{\nu-\mu}-\delta_{\mu \nu}
\end{equation}
where we defined $\hat{O}^{ee}_{\nu-\mu} \equiv N^{-1} \sum_n \alpha_{\mu,n}^* \alpha_{\nu,n} \hat{\sigma}^{ee}_n$. This allows to derive similar equations as in Section~\ref{supp:cumulant-collective} but for arbitrary geometries.

\subsection*{Three-dimensional arrays in free-space}
\begin{figure}[ht!]
    \centering
    \includegraphics[width = 1\columnwidth]{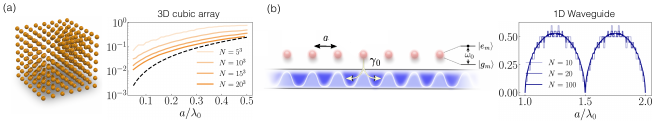}
    \caption{(a) The fraction of radiant modes for a three-dimensional cubic array of emitters with lattice spacing $a$, where radiant modes are defined as having decay rates $>\gamma_0/N$. The black dashed line indicates the value $(a/\lambda_0)^2$. The dipole moment matrix elements are chosen to be $\boldsymbol{d} = (1,i,0)^T$. (b) The radiant mode fraction for a chain of two-level emitters coupled via a one-dimensional, single-mode waveguide. Due to the periodic, infinite range interaction in Eq.~\ref{supp:wg}, a periodic modulation of the radiant mode fraction is observed ($k_0 = 2\pi/\lambda_0$). In the limiting case when $a=n\lambda_0/2$ ($n=1,2,3,\cdots$), all coherent couplings are zero, while the magnitude of all $\Gamma_{nm}$ is $\gamma_0$. The one-dimensional waveguide example shows, that at least half of the modes are subradiant for any spacing $a$, while for spacings close to multiples of $\lambda_0/2$, most modes becomes subradiant.}
    \label{fig:supp2}
\end{figure}

In the main text, we treat one- and two-dimensional geometries, and here a brief illustration of the radiant mode fraction for three-dimensional cubic arrays is shown in Fig.~\ref{fig:supp2}(a). We assume an ordered array of $N = N_\mathrm{1D} \times N_\mathrm{1D} \times N_\mathrm{1D} $ emitters with lattice spacing $a$ in all directions and define radiant modes, as having a decay rate $\Gamma_{{\mu}}>\gamma_0/N$, obtained either via Eq.~\ref{supp:rates}(a) or by numerical diagonalization of the decay matrix $\boldsymbol{\Gamma}$.
\subsection*{Waveguide quantum electrodynamics in one dimension}
The collective mode truncation can be applied to ordered emitter arrays coupled via one-dimensional reservoirs as well, because of the presence of collective dipole-dipole couplings and emerging super- and subradiant modes~\cite{asenjo2017exponential}.
In the case of a chain of two-level emitters at positions $\{x_n \}$ coupled via a single-mode, one-dimensional waveguide, the dipole-dipole couplings now read~\cite{sheremet2023waveguide,liedl2024observation}
\begin{equation} \label{supp:wg}
       J_{nm} - \frac{i\Gamma_{nm}}{2} = -\frac{i\gamma_0}{2} \ \mathrm{exp}\Big(i k_0 |{x_n}-x_m| \Big),
\end{equation}
which exhibits an infinite range, periodic modulation and where $k_0=\omega_0/c$ is the wavevector of the guided mode on resonance with the emitters.
In Fig.~\ref{fig:supp2}(b) shows, how the radiant mode fraction depends on the nearest-neighbor separation $a = |x_{n}-x_{n+1}|$, where we again define a radiant mode as having a decay rate $\Gamma_\mu>\gamma_0/N$. The collective decay rates are obtained via Eq.~\ref{supp:rates}(a), with the collective dipole couplings in Eq.~\ref{supp:wg}.

\begin{figure}[ht!]
    \centering
    \includegraphics[width = 1\columnwidth]{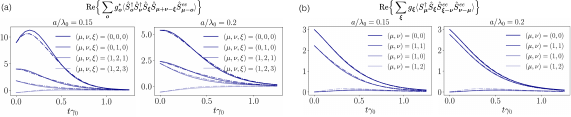}
    \caption{Comparison for the cumulant expansions performed for the two types of expectation values in Eq.~\ref{supp:expansion} and Eq.~\ref{supp:expansion2}. (a) Quasi-momenta combinations $(\mu,\nu,\xi)$ are shown for a ring geometry and $N=10$ emitters, with the largest contributions coming from modes with quasi-momenta close to $(\mu,\nu,\xi) = (0,0,0)$. (b) The same parameters but for the expectation value of the form as in Eq.~\ref{supp:expansion} with quasi-momenta $(\mu,\nu)$. In all plots: $\boldsymbol{d} = (1,i,0)^T$.}
    \label{fig:supp3}
\end{figure}
\section{Equations of motion in the collective basis} \label{supp:cumulant-collective}
We choose the set of collective operators $\{\hat{S}^\dagger_{\boldsymbol{\mu}},\hat{S}_{\boldsymbol{\mu}},\hat{S}^{ee}_{\boldsymbol{\mu}} \}$ to obtain all expectation values via  Eq.~\ref{heisenberg-collective}, necessary for computing the observables introduced in Section~\ref{section2} of the main text.
In order to obtain a closed set of equations, only correlations up to a certain order are considered, while all higher order correlations are expanded in terms of lower orders. This approximation can be computed in a systematic fashion, using the cumulant expansion~\cite{rubies2023characterizing,robicheaux2021beyond,sanchez2020cumulant}. The joint cumulant of an $N$-th order correlation $\langle \hat{O}_1...\hat{O}_N \rangle$ can be expressed as an alternate sum of products of their expectation values and is compactly written as~\cite{kubo1962generalized}
\begin{equation} \label{jointcumulant}
    \kappa(\hat{O}_1,...,\hat{O}_N) = \sum_\pi (|\pi|-1)!(-1)^{|\pi|-1} \prod_{B \in \pi} \Big\langle \prod_{i\in B} \hat{O}_i \Big\rangle
\end{equation}
where $\pi$ runs through the list of all partitions of $\{1, ..., N\}$, $B$ runs through the list of all blocks of the partition $\pi$ and $|\pi|$ is the number of parts in the partition. An approximation can be made by setting $\kappa(\hat{O}_1,...,\hat{O}_N) = 0 $, which allows to express $\langle \hat{O}_1...\hat{O}_N \rangle$ with correlations of order $N-1$ and lower. For instance, the expansion of expectation values involving three operators read
\begin{align}
    \langle \hat{O}_1 \hat{O}_2 \hat{O}_3 \rangle \approx \langle \hat{O}_1  \rangle \langle \hat{O}_2 \hat{O}_3  \rangle + \langle \hat{O}_2  \rangle \langle \hat{O}_1 \hat{O}_3  \rangle + \langle \hat{O}_3  \rangle \langle \hat{O}_1 \hat{O}_2  \rangle - 2 \langle \hat{O}_1  \rangle \langle \hat{O}_2   \rangle \langle \hat{O}_3 \rangle.
\end{align}
This introduces an error, since certain correlation are neglected and generally this error increases for smaller array spacings $a$, because correlations induced by strong dipole-dipole interactions increase to higher orders.
We note, that the influence of the coherent part with couplings $J_{nm}$ on the dissipative dynamics in Eq.~\ref{master} has been shown to be small~\cite{masson2024dicke}, at least for array spacings $a \gtrsim \lambda_0/10$. Nonetheless we include the coherent energy shifts and define the collective couplings in Eq.~\ref{collective-rates} as $g_{\boldsymbol{\mu}}= 2i J_{\boldsymbol{\mu}} + \Gamma_{\boldsymbol{\mu}}$.

Starting from the Heisenberg equations of motion in Eq.~\ref{heisenberg-collective} and including expectation values involving up to four operators, the closed set of equations read
    \begin{subequations} \label{supp:collective}
\begin{align} 
    \frac{d}{dt} \langle \hat{S}^{ee}_{\boldsymbol{0}} \rangle&= -\frac{1}{N} \sum_{\boldsymbol{\mu}} \Gamma_{\boldsymbol{\mu}} \langle \hat{S}^{\dagger}_{\boldsymbol{\mu}} \hat{S}_{\boldsymbol{\mu}} \rangle \\ \nonumber \\
\frac{d}{dt}\langle \hat{S}^{\dagger}_{\boldsymbol{\mu}} \hat{S}_{\boldsymbol{\mu}} \rangle&= -\Big(\Gamma_{\boldsymbol{\mu}} + \frac{1}{N}\sum_{\boldsymbol{\nu}} g_{\boldsymbol{\nu}} \Big) \langle \hat{S}^{\dagger}_{\boldsymbol{\mu}} \hat{S}_{\boldsymbol{\mu}} \rangle -\frac{1}{N} \sum_{\boldsymbol{\nu}} g^*_{\boldsymbol{\nu}} \langle \hat{S}^{\dagger}_{\boldsymbol{\nu}} \hat{S}_{\boldsymbol{\nu}} \rangle + \sum_{\boldsymbol{\nu}} \Big( g_{\boldsymbol{\nu}} \langle \hat{S}^{\dagger}_{\boldsymbol{\mu}} \hat{S}_{\boldsymbol{\nu}} \hat{S}^{ee}_{\boldsymbol{\nu-\mu}} \rangle + g^*_{\boldsymbol{\nu}} \langle \hat{S}^{\dagger}_{\boldsymbol{\nu}} \hat{S}_{\boldsymbol{\mu}} \hat{S}^{ee}_{\boldsymbol{\mu-\nu}} \rangle \Big) \\ \nonumber \\
\frac{d}{dt} \langle \hat{{S}}_{\boldsymbol{\mu}}^{ee} \hat{{S}}_{-\boldsymbol{\mu}}^{ee} \rangle&= \frac{1}{2N^2}\sum_{\boldsymbol{\nu}} g_{\boldsymbol{\nu}} \Big(\langle \hat{S}^\dagger_{\boldsymbol{\nu+\mu}} \hat{S}_{\boldsymbol{\nu+\mu}} \rangle + \langle \hat{S}^\dagger_{\boldsymbol{\nu-\mu}} \hat{S}_{\boldsymbol{\nu-\mu}} \rangle \Big)  \\
&-\frac{1}{2N}\sum_{\boldsymbol{\nu}} \Big( g_{\boldsymbol{\nu}} \langle \hat{S}^\dagger_{\boldsymbol{\nu+\mu}} \hat{S}_{\boldsymbol{\nu}} \hat{S}^{ee}_{-\boldsymbol{\mu}} \rangle + g_{\boldsymbol{\nu}} \langle \hat{S}^\dagger_{\boldsymbol{\nu-\mu}} \hat{S}_{\boldsymbol{\nu}} \hat{S}^{ee}_{\boldsymbol{\mu}} \rangle + g^*_{\boldsymbol{\nu}}\langle \hat{S}^\dagger_{\boldsymbol{\nu}} \hat{S}_{\boldsymbol{\nu-\mu}} \hat{S}^{ee}_{-\boldsymbol{\mu}} \rangle + g^*_{\boldsymbol{\nu}} \langle S^\dagger_{\boldsymbol{\nu}} \hat{S}_{\boldsymbol{\nu+\mu}} \hat{S}^{ee}_{\boldsymbol{\mu}} \rangle \Big) \nonumber \\ \nonumber \\
\frac{d}{dt} \langle \hat{{S}}^\dagger_{\boldsymbol{\mu}} \hat{{S}}_{\boldsymbol{\nu}} \hat{{S}}_{\boldsymbol{\nu-\mu}}^{ee} \rangle&= -\frac{1}{2N}\sum_{\boldsymbol{\xi}}  \Big(g_{\boldsymbol{\xi}}^*\langle S^\dagger_{\boldsymbol{\xi}} S^\dagger_{\boldsymbol{\mu}} \hat{S}_{\boldsymbol{\nu}} \hat{S}_{\boldsymbol{\xi+\mu-\nu}} \rangle + g_{\boldsymbol{\xi}} \langle S^\dagger_{\boldsymbol{\mu}} S^\dagger_{\boldsymbol{\xi+\mu-\nu}} \hat{S}_{\boldsymbol{\nu}} \hat{S}_{\boldsymbol{\xi}} \rangle \Big) +\Big(\frac{g_{\boldsymbol{\nu}}}{2N}-\frac{g_{\boldsymbol{\mu}}}{2N} \Big) \langle S^\dagger_{\boldsymbol{\mu}} \hat{S}_{\boldsymbol{\mu}} \rangle  \nonumber \\
&+ \frac{1}{N} \sum_{\boldsymbol{\xi}} \Big( g_{\boldsymbol{\xi}}\langle S^\dagger_{\boldsymbol{\mu}} \hat{S}_{\boldsymbol{\xi}} \hat{S}_{\boldsymbol{\xi-\mu}}^{ee}\rangle -g_{\boldsymbol{\xi}}^*\langle S^\dagger_{\boldsymbol{\xi}} \hat{S}_{\boldsymbol{\nu-\mu+\xi}} \hat{S}_{\boldsymbol{\nu-\mu}}^{ee}\rangle - g_{\boldsymbol{\xi}} \langle S^\dagger_{\boldsymbol{\mu}} \hat{S}_{\boldsymbol{\xi+\mu-\nu}} \hat{S}_{\boldsymbol{\xi-\mu}}^{ee}\rangle \Big) \nonumber \\
&-\Big(\frac{g^*_{\boldsymbol{\mu}}}{2}+\frac{g_{\boldsymbol{\nu}}}{2}+\sum_{\boldsymbol{\xi}} \frac{g_{\boldsymbol{\xi}}}{N} \Big) \langle S^\dagger_{\boldsymbol{\mu}} \hat{S}_{\boldsymbol{\nu}} \hat{S}_{\boldsymbol{\nu-\mu}}^{ee}\rangle + \sum_{\boldsymbol{\xi}} \Big( \underbrace{ g^*_{\boldsymbol{\xi}}\langle S^\dagger_{\boldsymbol{\xi}} \hat{S}_{\boldsymbol{\nu}} \hat{S}^{ee}_{\boldsymbol{\mu-\xi}} \hat{S}^{ee}_{\boldsymbol{\nu-\mu}} \rangle  + g_{\boldsymbol{\xi}} \langle  S^\dagger_{\boldsymbol{\mu}} \hat{S}_{\boldsymbol{\xi}} \hat{S}^{ee}_{\boldsymbol{\xi-\nu}} \hat{S}^{ee}_{\boldsymbol{\nu-\mu}}\rangle}_{\text{ cumulant expansion}} \Big) \nonumber \\ \nonumber \\
\frac{d}{dt} \langle \hat{{S}}^\dagger_{\boldsymbol{\mu}} \hat{{S}}_{\boldsymbol{\nu}}^\dagger \hat{{S}}_{\boldsymbol{\xi}} \hat{{S}}_{\boldsymbol{\mu+\nu-\xi}} \rangle &= -\Big(\sum_{\boldsymbol{o}} \frac{2g_{\boldsymbol{o}}}{N}+\frac{1}{2}(g_{\boldsymbol{\mu}}^*+g_{\boldsymbol{\nu}}^*+g_{\boldsymbol{\xi}}+g_{\boldsymbol{\mu+\nu-\xi}}) \Big) \langle S^{\dagger}_{\boldsymbol{\mu}} S^{\dagger}_{\boldsymbol{\nu}} \hat{S}_{\boldsymbol{\xi}} \hat{S}_{\boldsymbol{\mu+\nu-\xi}}\rangle  \\
&- \frac{1}{N}\sum_{\boldsymbol{o}} \Big(
g_{\boldsymbol{o}}^* \langle S^\dagger_{\boldsymbol{o}} S^\dagger_{\boldsymbol{\mu}} \hat{S}_{\boldsymbol{\xi-\nu+o}} \hat{S}_{\boldsymbol{\mu+\nu-\xi}} \rangle 
+g_{\boldsymbol{o}}^* \langle  S^\dagger_{\boldsymbol{o}} S^\dagger_{\boldsymbol{\mu}} \hat{S}_{\boldsymbol{\xi}} \hat{S}_{\boldsymbol{\mu-\xi+o}} \rangle  
- g_{\boldsymbol{o}}^* \langle S^\dagger_{\boldsymbol{o}} S^\dagger_{\boldsymbol{\mu+\nu-o}} \hat{S}_{\boldsymbol{\xi}} \hat{S}_{\boldsymbol{\mu+\nu-\xi}} \rangle \nonumber \\
&+ g_{\boldsymbol{o}}^* \langle S^\dagger_{\boldsymbol{o}} S^\dagger_{\boldsymbol{\nu}} \hat{S}_{\boldsymbol{\xi-\mu+o}} \hat{S}_{\boldsymbol{\mu+\nu-\xi}} \rangle 
+ g_{\boldsymbol{o}}^* \langle S^\dagger_{\boldsymbol{o}} S^\dagger_{\boldsymbol{\nu}} \hat{S}_{\boldsymbol{\xi}} \hat{S}_{\boldsymbol{o+\nu-\xi}}\rangle
+g_{\boldsymbol{o}} \langle S^\dagger_{\boldsymbol{\mu}} S^\dagger_{\boldsymbol{\nu}} \hat{S}_{\boldsymbol{\mu+\nu-o}} \hat{S}_{\boldsymbol{o}} \rangle \Big) \nonumber \\
&+ \sum_{\boldsymbol{o}} \Big( \underbrace{ g_{\boldsymbol{o}}^* \langle \hat{S}^\dagger_{\boldsymbol{o}} \hat{S}^\dagger_{\boldsymbol{\nu}} \hat{S}_{\boldsymbol{\xi}} \hat{S}_{\boldsymbol{\mu+\nu-\xi}} \hat{S}^{ee}_{\boldsymbol{\mu-o}} \rangle  + g_{\boldsymbol{o}}^* \langle \hat{S}^\dagger_{\boldsymbol{\mu}} \hat{S}^\dagger_{\boldsymbol{\nu}} \hat{S}_{\boldsymbol{\mu+\nu-\xi}} \hat{S}_{\boldsymbol{o}} \hat{S}^{ee}_{\boldsymbol{o-\xi}} \rangle+ g_{\boldsymbol{o}} \langle \hat{S}^\dagger_{\boldsymbol{\mu}} \hat{S}^\dagger_{\boldsymbol{\nu}} \hat{S}_{\boldsymbol{\xi}} \hat{S}_{\boldsymbol{o}} \hat{S}^{ee}_{\boldsymbol{o+\xi-\mu-\nu}} \rangle }_{\text{ cumulant expansion}} \nonumber \\
&+ \underbrace{ g_{\boldsymbol{o}} \langle \hat{S}^\dagger_{\boldsymbol{o}} \hat{S}^\dagger_{\boldsymbol{\mu}} \hat{S}_{\boldsymbol{\xi}} \hat{S}_{\boldsymbol{\mu+\nu-\xi}}  \hat{S}^{ee}_{\boldsymbol{\nu-o}} \rangle}_{\text{ cumulant expansion}} \Big) \nonumber 
\end{align}
\end{subequations}
The expansions of the expectation values highlighted in Eqs.~\ref{supp:collective} are given by
    \begin{subequations} \label{supp:expansion}
\begin{align}
   \sum_{\boldsymbol{\xi}} g_{\boldsymbol{\xi}}  \langle  S^\dagger_{\boldsymbol{\mu}} \hat{S}_{\boldsymbol{\xi}} \hat{S}^{ee}_{\boldsymbol{\xi-\nu}} \hat{S}^{ee}_{\boldsymbol{\nu-\mu}}\rangle &\approx \delta_{\boldsymbol{\mu \nu}} \langle \hat{S}^{ee}_{\boldsymbol{0}} \rangle \textstyle \sum_{\boldsymbol{\xi}} g_{\boldsymbol{\xi}} \langle \hat{S}^\dagger_{\boldsymbol{\mu}} \hat{S}_{\boldsymbol{\xi}} \hat{S}^{ee}_{\boldsymbol{\xi-\mu}} \rangle + g_{\boldsymbol{\nu}} \langle \hat{S}^{ee}_{\boldsymbol{0}} \rangle  \langle \hat{S}^\dagger_{\boldsymbol{\mu}} \hat{S}_{\boldsymbol{\nu}} \hat{S}^{ee}_{\boldsymbol{\nu-\mu}} \rangle \\
  &+g_{\boldsymbol{\mu}} \langle \hat{S}^\dagger_{\boldsymbol{\mu}} \hat{S}_{\boldsymbol{\mu}} \rangle \langle \hat{S}^{ee}_{\boldsymbol{\mu-\nu}} \hat{S}^{ee}_{\boldsymbol{\nu-\mu}} \rangle -2g_{\boldsymbol{\mu}} \langle \hat{S}^{ee}_{\boldsymbol{0}} \rangle^2 \langle \hat{S}^\dagger_{\boldsymbol{\mu}} \hat{S}_{\boldsymbol{\mu}} \rangle \delta_{\boldsymbol{\mu \nu}}. \nonumber \\ \nonumber \\
    \sum_{\boldsymbol{\xi}} g_{\boldsymbol{\xi}}^* \langle S^\dagger_{\boldsymbol{\xi}} \hat{S}_{\boldsymbol{\nu}} \hat{S}^{ee}_{\boldsymbol{\mu-\xi}} \hat{S}^{ee}_{\boldsymbol{\nu-\mu}} \rangle &\approx \delta_{\boldsymbol{\mu \nu}} \langle \hat{S}^{ee}_{\boldsymbol{0}} \rangle \textstyle \sum_{\boldsymbol{\xi}} g^*_{\boldsymbol{\xi}} \langle \hat{S}^\dagger_{\boldsymbol{\xi}} \hat{S}_{{\boldsymbol{\mu}}} \hat{S}^{ee}_{\boldsymbol{\mu-\xi}} \rangle + g^*_{\boldsymbol{\mu}} \langle \hat{S}^{ee}_{\boldsymbol{0}} \rangle  \langle \hat{S}^\dagger_{\boldsymbol{\mu}} \hat{S}_{{\boldsymbol{\nu}}} \hat{S}^{ee}_{\boldsymbol{\nu-\mu}} \rangle \\
  &+g_{\boldsymbol{\nu}}^* \langle \hat{S}^\dagger_{\boldsymbol{\nu}} \hat{S}_{\boldsymbol{\nu}} \rangle \langle \hat{S}^{ee}_{\boldsymbol{\mu-\nu}} \hat{S}^{ee}_{\boldsymbol{\nu-\mu}} \rangle -2 g_{\boldsymbol{\mu}}^* \langle \hat{S}^{ee}_{\boldsymbol{0}} \rangle^2 \langle \hat{S}^\dagger_{\boldsymbol{\mu}} \hat{S}_{\boldsymbol{\mu}} \rangle \delta_{\boldsymbol{\mu \nu}}. \nonumber
\end{align}
\end{subequations}

    \begin{subequations} \label{supp:expansion2}
\begin{align}
 \sum_{\boldsymbol{o}} g_{\boldsymbol{o}}^* \langle \hat{S}^\dagger_{\boldsymbol{o}} \hat{S}^\dagger_{\boldsymbol{\nu}} \hat{S}_{\boldsymbol{\xi}} \hat{S}_{\boldsymbol{\mu+\nu-\xi}} \hat{S}^{ee}_{\boldsymbol{\mu-o}} \rangle &\approx g^*_{\boldsymbol{\mu}} \langle \hat{S}^\dagger_{\boldsymbol{\mu}} \hat{S}^\dagger_{\boldsymbol{\nu}} \hat{S}_{\boldsymbol{\xi}} \hat{S}_{\boldsymbol{\mu+\nu-\xi}}  \rangle \langle \hat{S}^{ee}_{\boldsymbol{0}} \rangle - 2 g_{\boldsymbol{\mu}}^* (\delta_{\boldsymbol{\mu \xi}}+\delta_{\boldsymbol{\nu \xi}}) \langle \hat{S}^\dagger_{\boldsymbol{\mu}} \hat{S}_{\boldsymbol{\mu}} \rangle  \langle \hat{S}^\dagger_{\boldsymbol{\nu}} \hat{S}_{\boldsymbol{\nu}} \rangle \langle \hat{S}^{ee}_{\boldsymbol{0}} \rangle 
 \\
&+\sum_{\boldsymbol{o}} g_{\boldsymbol{o}}^* (\delta_{\boldsymbol{\mu \xi}}+\delta_{\boldsymbol{\nu \xi}}) \langle \hat{S}^\dagger_{\boldsymbol{\nu}} \hat{S}_{\boldsymbol{\nu}} \rangle   \langle \hat{S}^\dagger_{\boldsymbol{o}} \hat{S}_{\boldsymbol{\mu}} \hat{S}^{ee}_{\boldsymbol{\mu-o}} \rangle + g^*_{\boldsymbol{\xi}} \langle \hat{S}^\dagger_{\boldsymbol{\xi}} \hat{S}_{\boldsymbol{\xi}} \rangle   \langle \hat{S}^\dagger_{\boldsymbol{\nu}} \hat{S}_{\boldsymbol{\mu+\nu-\xi}} \hat{S}^{ee}_{\boldsymbol{\mu-\xi}} \rangle \nonumber \\
&+ g_{\boldsymbol{\mu+\nu-\xi}}^*  \langle \hat{S}^\dagger_{\boldsymbol{\mu+\nu-\xi}} \hat{S}_{\boldsymbol{\mu+\nu-\xi}}\rangle   \langle \hat{S}^\dagger_{\boldsymbol{\nu}} \hat{S}_{\boldsymbol{\xi}} \hat{S}^{ee}_{\boldsymbol{\xi-\nu}} \rangle \nonumber \\ \nonumber \\
\sum_{\boldsymbol{o}} g_{\boldsymbol{o}}^* \langle \hat{S}^\dagger_{\boldsymbol{\mu}} \hat{S}^\dagger_{\boldsymbol{\nu}} \hat{S}_{\boldsymbol{\mu+\nu-\xi}}  \hat{S}_{\boldsymbol{o}} \hat{S}^{ee}_{\boldsymbol{o-\xi}} \rangle &\approx g^*_{\boldsymbol{\xi}} \langle \hat{S}^\dagger_{\boldsymbol{\mu}} \hat{S}^\dagger_{\boldsymbol{\nu}} \hat{S}_{\boldsymbol{\mu+\nu-\xi}} \hat{S}_{\boldsymbol{\xi}} \rangle \langle \hat{S}^{ee}_{\boldsymbol{0}} \rangle - 2 g^*_{\boldsymbol{\xi}} (\delta_{\boldsymbol{\mu \xi}}+\delta_{\boldsymbol{\nu \xi}}) \langle \hat{S}^\dagger_{\boldsymbol{\mu}} \hat{S}_{\boldsymbol{\mu}} \rangle  \langle \hat{S}^\dagger_{\boldsymbol{\nu}} \hat{S}_{\boldsymbol{\nu}} \rangle \langle \hat{S}^{ee}_{\boldsymbol{0}} \rangle 
 \\
&+\sum_{\boldsymbol{o}} g^*_{\boldsymbol{o}} (\delta_{\boldsymbol{\mu \xi}}+\delta_{\boldsymbol{\nu \xi}}) \langle \hat{S}^\dagger_{\boldsymbol{\xi}} \hat{S}_{\boldsymbol{\xi}} \rangle   \langle \hat{S}^\dagger_{\boldsymbol{\xi}} \hat{S}_{\boldsymbol{o}} \hat{S}^{ee}_{\boldsymbol{o-\xi}} \rangle + g^*_{\boldsymbol{\mu}} \langle \hat{S}^\dagger_{\boldsymbol{\mu}} \hat{S}_{\boldsymbol{\mu}} \rangle   \langle \hat{S}^\dagger_{\boldsymbol{\nu}} \hat{S}_{\boldsymbol{\xi}} \hat{S}^{ee}_{\boldsymbol{\xi-\nu}} \rangle \nonumber \\
&+ g_{\boldsymbol{\nu}}^*  \langle \hat{S}^\dagger_{\boldsymbol{\nu}} \hat{S}_{\boldsymbol{\nu}} \rangle   \langle \hat{S}^\dagger_{\boldsymbol{\mu}} \hat{S}_{\boldsymbol{\xi}} \hat{S}^{ee}_{\boldsymbol{\xi-\mu}} \rangle \nonumber \\ \nonumber \\
\sum_{\boldsymbol{o}} g_{\boldsymbol{o}} \langle \hat{S}^\dagger_{\boldsymbol{\mu}} \hat{S}^\dagger_{\boldsymbol{\nu}} \hat{S}_{\boldsymbol{\xi}} \hat{S}_{\boldsymbol{o}} \hat{S}^{ee}_{\boldsymbol{o+\xi-\mu-\nu}} \rangle &\approx g_{\boldsymbol{\mu+\nu-\xi}} \langle \hat{S}^\dagger_{\boldsymbol{\mu}} \hat{S}^\dagger_{\boldsymbol{\nu}} \hat{S}_{\boldsymbol{\xi}} \hat{S}_{\boldsymbol{\mu+\nu-\xi}} \rangle \langle \hat{S}^{ee}_{\boldsymbol{0}} \rangle - 2 g_{\boldsymbol{\mu+\nu-\xi}} (\delta_{\boldsymbol{\mu \xi}}+\delta_{\boldsymbol{\nu \xi}}) \langle \hat{S}^\dagger_{\boldsymbol{\mu}} \hat{S}_{\boldsymbol{\mu}} \rangle  \langle \hat{S}^\dagger_{\boldsymbol{\nu}} \hat{S}_{\boldsymbol{\nu}} \rangle \langle \hat{S}^{ee}_{\boldsymbol{0}} \rangle  
 \\
&+\sum_{\boldsymbol{o}} g_{\boldsymbol{o}} (\delta_{\boldsymbol{\mu \xi}}+\delta_{\boldsymbol{\nu \xi}}) \langle \hat{S}^\dagger_{\boldsymbol{\xi}} \hat{S}_{\boldsymbol{\xi}} \rangle   \langle \hat{S}^\dagger_{\boldsymbol{\nu}} \hat{S}_{\boldsymbol{o}} \hat{S}^{ee}_{\boldsymbol{o-\nu}} \rangle + g_{\boldsymbol{\mu}} \langle \hat{S}^\dagger_{\boldsymbol{\mu}} \hat{S}_{\boldsymbol{\mu}} \rangle   \langle \hat{S}^\dagger_{\boldsymbol{\nu}} \hat{S}_{\boldsymbol{\xi}} \hat{S}^{ee}_{\boldsymbol{\xi-\nu}} \rangle \nonumber \\
&+ g_{\boldsymbol{\nu}}  \langle \hat{S}^\dagger_{\boldsymbol{\nu}} \hat{S}_{\boldsymbol{\nu}} \rangle   \langle \hat{S}^\dagger_{\boldsymbol{\mu}} \hat{S}_{\boldsymbol{\xi}} \hat{S}^{ee}_{\boldsymbol{\xi-\mu}} \rangle \nonumber \\ \nonumber \\
\sum_{\boldsymbol{o}} g_{\boldsymbol{o}} \langle \hat{S}^\dagger_{\boldsymbol{o}} \hat{S}^\dagger_{\boldsymbol{\mu}} \hat{S}_{\boldsymbol{\xi}} \hat{S}_{\boldsymbol{\mu+\nu-\xi}}\hat{S}^{ee}_{\boldsymbol{\nu-o}} \rangle &\approx g_{\boldsymbol{\nu}} \langle \hat{S}^\dagger_{\boldsymbol{\nu}} \hat{S}^\dagger_{\boldsymbol{\mu}} \hat{S}_{\boldsymbol{\xi}} \hat{S}_{\boldsymbol{\mu+\nu-\xi}} \rangle \langle \hat{S}^{ee}_{\boldsymbol{0}} \rangle - 2 g_{\boldsymbol{\nu}} (\delta_{\boldsymbol{\mu \xi}}+\delta_{\boldsymbol{\nu \xi}}) \langle \hat{S}^\dagger_{\boldsymbol{\mu}} \hat{S}_{\boldsymbol{\mu}} \rangle  \langle \hat{S}^\dagger_{\boldsymbol{\nu}} \hat{S}_{\boldsymbol{\nu}} \rangle \langle \hat{S}^{ee}_{\boldsymbol{0}} \rangle 
 \\
&+\sum_o g_{\boldsymbol{o}}(\delta_{\boldsymbol{\mu \xi}}+\delta_{\boldsymbol{\nu \xi}}) \langle \hat{S}^\dagger_{\boldsymbol{\mu}} \hat{S}_{\boldsymbol{\mu}} \rangle \langle \hat{S}^\dagger_{\boldsymbol{o}} \hat{S}_{\boldsymbol{\nu}} \hat{S}^{ee}_{\boldsymbol{\nu-o}} \rangle + g_{\boldsymbol{\xi}} \langle \hat{S}^\dagger_{\boldsymbol{\xi}} \hat{S}_{\boldsymbol{\xi}} \rangle   \langle \hat{S}^\dagger_{\boldsymbol{\mu}} \hat{S}_{\boldsymbol{\mu+\nu-\xi}} \hat{S}^{ee}_{\boldsymbol{\nu-\xi}} \rangle \nonumber \\
&+ g_{\boldsymbol{\mu+\nu-\xi}}  \langle \hat{S}^\dagger_{\boldsymbol{\mu+\nu-\xi}} \hat{S}_{\boldsymbol{\mu+\nu-\xi}} \rangle   \langle \hat{S}^\dagger_{\boldsymbol{\mu}} \hat{S}_{\boldsymbol{\xi}} \hat{S}^{ee}_{\boldsymbol{\xi-\nu}} \rangle, \nonumber 
\end{align}
\end{subequations}
where we defined $\boldsymbol{0} = (0,0)$ generally of two-dimensional arrays.
In Fig.~\ref{fig:supp3} we show a comparison of the left- and right-hand side of the expansions in Eqs.~\ref{supp:expansion} and Eqs.~\ref{supp:expansion2} via the master equation for emitters in a ring geometry. Excellent agreement is observed for the expansions in Eq.~\ref{supp:expansion}(a) and Eq.~\ref{supp:expansion}(b), thus we neglect the equation of motion of the expectation values of the form $\langle  S^\dagger_{\boldsymbol{\mu}} \hat{S}_{\boldsymbol{\xi}} \hat{S}^{ee}_{\boldsymbol{\xi-\nu}} \hat{S}^{ee}_{\boldsymbol{\nu-\mu}}\rangle$, as they would involve multiple higher order correlations, rendering the overall system of equations much more complex.
The largest errors stem from the expansions in Eqs.~\ref{supp:expansion2}, in particular for the expectation value with quasi-momenta $(\boldsymbol{\mu,\nu,\xi})=(\boldsymbol{0,0,0})$.

\section{Cumulant expansion for individual spin operators} \label{sup:cumulant}
In the main text, we compare the collective mode truncation, with equations of motion for individual spin operators based on second- and third-order cumulant expansions. Starting from the Heisenberg equations of motion
\begin{equation}
    \frac{d}{dt} \langle \hat{O} \rangle =  i \langle [\hat{H},\hat{O}]\rangle 
    - \sum_{i,j=1}^N \frac{\Gamma_{ij}}{2} \Big(\langle \hat{\sigma}^\dagger_i [\hat{\sigma}_j,\hat{O}]\rangle - \langle [\hat{\sigma}^\dagger_i,\hat{O}] \hat{\sigma}_j \rangle \Big),
\end{equation}
and using the commutation relations for individual spin operators
\begin{subequations}
\begin{align}
    [\hat{\sigma}_i^\dagger,\hat{\sigma}_{j} ] &= (2 \hat{\sigma}_{i}^{ee}  - \hat{\mathds{1}}) \delta_{ij}, \\
[\hat{\sigma}^\dagger_i,\hat{\sigma}_{j}^{ee} ]  &= - \hat{\sigma}^\dagger_{i} \delta_{ij}, \\
[\hat{\sigma}_i,\hat{\sigma}_{j}^{ee} ]  &= \hat{\sigma}_{i} \delta_{ij},
\end{align}
\end{subequations}
a closed set of equations can be derived for the set of $3N$ spin operators $\{\hat{\sigma}^\dagger_i,\hat{\sigma}_i,\hat{\sigma}^{ee}_i \}$. Similar equations can be obtained with the set of Pauli spin operators $\{\hat{\sigma}^x_i,\hat{\sigma}^y_i,\hat{\sigma}^z_i \}$~\cite{kramer2015generalized}.

\subsection*{Second-order cumulant expansion}
The cumulant expansions up to second order involve the correlations $\langle \hat{\sigma}_i^{ee} \rangle$, $\langle \hat{\sigma}_i^\dagger \hat{\sigma}_j \rangle$ and $\langle \hat{\sigma}_i^{ee} \hat{\sigma}_j^{ee} \rangle$ for $i\neq j$~\cite{sanchez2020cumulant,kira2008cluster}. We note, that because of the initial product state in Eq.~\ref{initial-state}, the correlations $\langle \hat{\sigma}_i \rangle$ and $\langle \hat{\sigma}_i \hat{\sigma}_j^{ee} \rangle$ remain zero at all times~\cite{rubies2023characterizing,masson2024dicke}. The equations read

    \begin{subequations}
\begin{align}
        \frac{d}{dt} \langle \hat{\sigma}^{ee}_i \rangle &= -\gamma_0 \langle \hat{\sigma}^{ee}_i \rangle + \sum_{n\neq i} \Big\{ \Big(iJ_{ni}-\frac{\Gamma_{ni}}{2} \Big)\langle \hat{\sigma}^{\dagger}_n \hat{\sigma}_i \rangle + \Big(-iJ_{in}-\frac{\Gamma_{in}}{2} \Big)\langle \hat{\sigma}^{\dagger}_i \hat{\sigma}_n \rangle   \Big\}, \\
        \frac{d}{dt} \langle \hat{\sigma}^{\dagger}_i \hat{\sigma}_j \rangle &= -\gamma_0 \langle \hat{\sigma}^{\dagger}_i \hat{\sigma}_j \rangle + \frac{\Gamma_{ij}}{2} (4 \langle \hat{\sigma}^{ee}_i \hat{\sigma}^{ee}_j \rangle - \langle \hat{\sigma}^{ee}_i \rangle -\langle \hat{\sigma}^{ee}_j \rangle) + i J_{ji}(\langle \hat{\sigma}^{ee}_j \rangle-\langle \hat{\sigma}^{ee}_i \rangle) \nonumber \\
        &+ \sum_{n\neq i,j} \Big\{ \Big(i J_{jn}+\frac{\Gamma_{jn}}{2} \Big) \langle \hat{\sigma}^{\dagger}_i \hat{\sigma}_n \rangle (2\langle \hat{\sigma}_j^{ee}\rangle-1)+ \Big(-i J_{in}+\frac{\Gamma_{in}}{2} \Big) \langle \hat{\sigma}^{\dagger}_n \hat{\sigma}_j \rangle (2\langle \hat{\sigma}_i^{ee}\rangle-1) \Big\}, \\
        \frac{d}{dt} \langle \hat{\sigma}^{ee}_i  \hat{\sigma}^{ee}_j \rangle &= -2\gamma_0 \langle \hat{\sigma}^{ee}_i  \hat{\sigma}^{ee}_j \rangle + \sum_{n\neq i,j} \Big\{ \Big(iJ_{nj}-\frac{\Gamma_{nj}}{2} \Big)\langle \hat{\sigma}^{ee}_i\rangle \langle \hat{\sigma}^{\dagger}_n \hat{\sigma}_j \rangle + \Big(-iJ_{jn}-\frac{\Gamma_{jn}}{2} \Big) \langle \hat{\sigma}^{ee}_i\rangle \langle \hat{\sigma}^{\dagger}_j \hat{\sigma}_n \rangle \nonumber \\
        &+ \Big(iJ_{ni}-\frac{\Gamma_{ni}}{2} \Big) \langle \hat{\sigma}^{ee}_j\rangle \langle \hat{\sigma}^{\dagger}_n \hat{\sigma}_i \rangle + \Big(-iJ_{ni}-\frac{\Gamma_{ni}}{2} \Big) \langle \hat{\sigma}^{ee}_j\rangle \langle \hat{\sigma}^{\dagger}_i \hat{\sigma}_n \rangle\Big\},
        \end{align}
\end{subequations}
and the initial expectation values at $t=0$ are given by $\langle \hat{\sigma}_i^{ee} \rangle = 1$, $\langle \sigma^\dagger_i \hat{\sigma}_j \rangle = 0$ and $\langle \sigma^{ee}_i \hat{\sigma}^{ee}_j \rangle = 1$ for $i\neq j$.

\subsection*{Third-order cumulant expansion}
To obtain a closed set of equations up to third order, the time evolution for the following set of expectation values has to be evaluated,
\begin{equation}
    \Big\{ \langle \hat{\sigma}^{ee}_i \rangle,  \langle \hat{\sigma}^{\dagger}_i \hat{\sigma}_j \rangle, \langle \hat{\sigma}^{ee}_i \hat{\sigma}^{ee}_j \rangle, \langle \hat{\sigma}^{ee}_i \hat{\sigma}^{ee}_j \hat{\sigma}^{ee}_k \rangle , \langle \hat{\sigma}^{\dagger}_i \hat{\sigma}_j \hat{\sigma}^{ee}_k \rangle \Big\}
\end{equation}
with $i\neq j\neq k$ and the expectation values at $t=0$ are given by $\langle \hat{\sigma}_i^{ee} \rangle = 1$, $\langle \sigma^\dagger_i \hat{\sigma}_j \rangle = 0$, $\langle \sigma^{ee}_i \hat{\sigma}^{ee}_j \rangle = 1$, $\langle \sigma^{ee}_i \hat{\sigma}^{ee}_j \hat{\sigma}^{ee}_k\rangle = 1$ and $\langle \sigma^{\dagger}_i \hat{\sigma}_j \hat{\sigma}^{ee}_k\rangle = 0$. We refer to the appendix in Reference~\cite{rubies2023characterizing}, where the resulting equations are provided in full detail.

\section{Directional photon emission} \label{supp:directional}
The time-evolved observables from the main text are evaluated in the collective basis and involve the total emitted emission integrated over all space. Equivalently, the emitted photon field at any point in space can be calculated based on the electric field operator~\cite{holzinger2021nanoscale,asenjo2017exponential,moreno2019subradiance}
\begin{equation}
     \hat{E}(\boldsymbol{r}) = \frac{|\boldsymbol{d}|k_0^2} {\epsilon_0}  \sum_{\boldsymbol{n}}^N \textbf{G}(\boldsymbol{r}-\boldsymbol{r}_{\boldsymbol{n}},\omega_0) \hat{\sigma}_{\boldsymbol{n}} = \frac{|\boldsymbol{d}|k_0^2} {\epsilon_0}  \sum_{\boldsymbol{\mu}} \tilde{\textbf{G}}_{\boldsymbol{\mu}}(\boldsymbol{r}) \hat{S}_{\boldsymbol{\mu}},
\end{equation}
where $\tilde{\textbf{G}}_{\boldsymbol{\mu}}(\boldsymbol{r}) = N^{-1/2} \sum_{\boldsymbol{n}} \mathrm{exp} [i 2\pi/N_\mathrm{1D} \boldsymbol{\mu} \cdot \boldsymbol{n}] \ \textbf{G}(\boldsymbol{r},\omega_0)$ and $\boldsymbol{n} = (n_x,n_y)$ labels the emitter in two-dimensional arrays with $N = N_\mathrm{1D} \times N_\mathrm{1D}$, while we set $\boldsymbol{n} = (n_x,0)$ for one-dimensional arrays.

The directional photon emission at any point $\boldsymbol{r}$ can now be calculated based on the expectation values $\langle \hat{S}_{\boldsymbol{\mu}}^\dagger \hat{S}_{\boldsymbol{\mu}}\rangle$ as
\begin{align}
    P_\mathrm{out}(\boldsymbol{r},t) = \langle \hat{E}^\dagger(\boldsymbol{r}) \hat{E}(\boldsymbol{r}) \rangle = \sum_{\boldsymbol{\mu}} |\tilde{\textbf{G}}_{\boldsymbol{\mu}}(\boldsymbol{r})|^2 \langle \hat{S}_{\boldsymbol{\mu}}^\dagger \hat{S}_{\boldsymbol{\mu}}\rangle.
\end{align}
Similarly, the second order correlation $g^{(2)}(\tau =0)$ in Eq.~\ref{eq:g2} can be evaluated at any point in space via the field operators $\hat{E}(\boldsymbol{r})$~\cite{masson2020many,holzinger2021nanoscale}. 

The time-evolved excited state population for an indivudal emitter $n$ can be calculated based on the expectation values $\langle \hat{S}_{\boldsymbol{\mu}}^\dagger \hat{S}_{\boldsymbol{\mu}}\rangle$ in the collective basis via
\begin{equation}
    \langle \hat{\sigma}^{ee}_{\boldsymbol{n}} \rangle = \langle \hat{\sigma}^{\dagger}_{\boldsymbol{n}} \hat{\sigma}_{\boldsymbol{n}} \rangle = \frac{1}{N} \sum_{\boldsymbol{\mu}} \mathrm{exp} \Big( i \frac{2\pi}{N_\mathrm{1D}} \boldsymbol{\mu}\cdot \boldsymbol{n} \Big) \langle \hat{S}_{\boldsymbol{\mu}}^\dagger \hat{S}_{\boldsymbol{\mu}}\rangle,
\end{equation}
where the summation runs over all quasi-momenta $\boldsymbol{\mu} = (\mu_x,\mu_y)$ in two-dimensional arrays and $\boldsymbol{\mu} = (\mu_x,0)$ in one-dimensional arrays.

\section{Coherent and incoherent illumination} \label{driving}

Here, we briefly discuss the generalization of the collective mode truncation method to driven systems. A temporally incoherent (but spatially coherent) pumping process can be modeled by extending the master
equation in Eq.~\ref{master} by the term~\cite{moreno2022efficient}
\begin{equation}
    \mathcal{L}_\mathrm{inc}[\hat{\rho}] = \frac{R}{2}\Big(2 \hat{S}^\dagger_{\boldsymbol{0}} \hat{\rho} \hat{S}_{\boldsymbol{0}} - \hat{S}_{\boldsymbol{0}}  \hat{S}^\dagger_{\boldsymbol{0}} \hat{\rho} - \hat{\rho}  \hat{S}_{\boldsymbol{0}}  \hat{S}^\dagger_{\boldsymbol{0}} \Big),
\end{equation}
where $R$ is the pumping rate and we assume a perpendicularly propagating beam, driving only the mode with quasi-momentum $\boldsymbol{\mu}=\boldsymbol{0}$. The resulting equations of motion are not sigfnicantly altered after including this incoherent driving term as it will not add coherence to the system, thus expectation values such as $\langle \hat{S}_{\boldsymbol{\mu}}\rangle$ remain zero at all times.

Conversely, by adding a coherent laser drive to the system's Hamiltonian, via $\Delta_l \hat{S}^\dagger_{\boldsymbol{0}}\hat{S}_{\boldsymbol{0}}+ \Omega_0 (\hat{S}^\dagger_{\boldsymbol{0}}+\hat{S}_{\boldsymbol{0}})$, a build-up of coherences occurs leading to expectation values $\langle \hat{S}_{\boldsymbol{\mu}}\rangle \neq 0$. Here, $\Delta_l$ is the laser detuning with respect to the bare emitter transition frequency $\omega_0$ and $\Omega_0$ the coherent driving rate.
Consequently, the time evolution of expectation values $ \langle \hat{S}_{\boldsymbol{\mu}}\rangle, \langle \hat{S}_{\boldsymbol{\mu}} \hat{S}^{ee}_{\boldsymbol{\mu}}\rangle,\langle \hat{S}^\dagger_{\boldsymbol{\mu}} \hat{S}^{ee}_{\boldsymbol{-\mu}}\rangle, \cdots $ , have to be included in the equations of motion.




\end{document}